\documentclass[aps,twocolumn,showpacs,floatfix,amssymb]{revtex4}
\usepackage{graphicx}
\begin{document}

\title{On the theory of superconductivity in
the extended Hubbard model:\\ Spin-fluctuation pairing}
\author{Nikolay M. Plakida$^{1, 2}$ and Viktor S. Oudovenko$^{3}$}
 \affiliation{$^1$Joint Institute for Nuclear Research,
 141980 Dubna, Russia}
 \affiliation{$^2$Max-Planck-Institut f\"{u}r Physik komplexer
Systeme, D-01187, Dresden,  Germany}
 \affiliation{$^3$Rutgers University,  New Jersey 08854, USA }

\date{\today}
\begin{abstract}
A microscopic theory of superconductivity  in the extended Hubbard model which takes into
account  the intersite Coulomb repulsion and electron-phonon interaction is developed in
the limit of strong correlations. The Dyson equation for  normal and pair Green functions
expressed in  terms of the Hubbard operators is derived. The self-energy is obtained  in
the noncrossing approximation.  In the normal state, antiferromagnetic short-range
correlations result in the electronic spectrum with a narrow bandwidth. We calculate
superconducting $T_c$  by taking into account the pairing mediated by charge and spin
fluctuations and phonons.  We found the $d$-wave pairing with high-$T_c$ mediated by spin
fluctuations induced by the strong kinematic interaction for the Hubbard operators.
Contributions to the $d$-wave pairing coming from the intersite Coulomb repulsion and
phonons turned out to be small.
\end{abstract}


 \maketitle

%------------------------------------
\section{Introduction}
\label{sec:1}

Despite  intensive studies of  high-temperature superconductivity (HTSC) in  cuprates for
many years after the discovery of Bednorz and M\"uller~\cite{Bednorz86}, a commonly
accepted mechanism of HTSC is still lacking (see, e.g.~\cite{Schrieffer07,Plakida10}). A
good candidate from various proposed   mechanisms is based on a model of strongly
correlated electrons~\cite{Anderson87}. In the model, superconductivity occurs at finite
doping in the resonating valence bond state (RVB) due to the antiferromagnetic (AF)
superexchange in the $t$--$J$ model. A possibility of HTSC mediated  by AF spin
fluctuations as a ``glue'' for superconducting pairing  was also
considered~\cite{Scalapino95}, mostly within phenomenological spin-fermion models (see,
e.g.,~\cite{Monthoux94b,Moriya00,Chubukov04,Abanov08}, and references therein).
\par
Recent studies of  spin-excitations  by magnetic inelastic neutron scattering (INS) and
the electronic spectrum by angle-resolved photoemission spectroscopy  (ARPES)  have
revealed an important role of AF spin excitations in the ``kink'' phenomenon and the
$d$-wave pairing in cuprates (see, e.g.,~\cite{Kordyuk10} and references therein). In
particular, in Ref.~\cite{Dahm09} using INS and  ARPES studies on the same
YBa$_2$Cu$_3$O$_{6.6}$ (YBCO$_{6.6}$) crystal,  an estimation for superconducting $T_c
\sim 150$~K was found. The main argument against the spin-fluctuation pairing, the weak
intensity of spin fluctuations at the optimal doping seen in  INS
experiments~\cite{Bourges98}, was dismissed in the recent resonant inelastic x-ray
scattering experiments~\cite{LeTacon11}. In a large family of cuprate superconductors,
paramagnon AF excitations with the dispersion and spectral weight similar to those of
magnons in undoped cuprates were observed. Using the magnetic spectrum found in the
YBCO$_7$ crystal, superconductivity  with  $T_c = 100-200$~K was predicted. Thus, spin
fluctuations have sufficient strength to mediate HTSC in cuprates and to explain various
physical properties of cuprate materials as, e.g., the optical
conductivity~\cite{Vladimirov12}. Therefore, it can  be suggested that the alternative
mechanism based on the conventional electron-phonon interaction (EPI) (see,
e.g.,~\cite{Kulic04,Maksimov10})  plays a secondary role in the cuprate superconductors.
\par
Recently, in Ref.~\cite{Raghu10} using the renormalization group (RG) method an
asymptotically exact solution for the $d$-wave pairing was found in the conventional
Hubbard model~\cite{Hubbard63} in the weak correlation limit, $U \ll t$. However, as was
pointed out later in Ref.~\cite{Alexandrov11},  a contribution from the repulsive
well-screened weak Coulomb interaction (CI) in the first order strongly suppresses the
pairing induced by the contributions of higher orders, and a possibility for
superconductivity ``from repulsion'' was questioned. At the same time, in
Ref.~\cite{KaganM11} it was shown that the $p$-wave superconductivity exists  in the
electronic gas at low density with a strong repulsion $U$ and a relatively strong Coulomb
intersite interaction $V_{ij}$ (see, also~\cite{Efremov00} and references therein). Later
on, in Ref.~\cite{Raghu12} RG studies of the extended Hubbard model with the intersite
interaction  have shown that superconducting pairing of various symmetries, extended
$s$-, $p$-, and $d$-wave types, can occur depending on the electron concentration and the
intersite interaction $V_{ij}$. However, in these investigations the Fermi-liquid model
in the weak correlation limit was used. To study superconductivity in cuprates,  the
Mott-Hubbard (more accurately, charge-transfer) doped insulators,  a theory of strongly
correlated electronic systems should be used (for reviews see~\cite{Fulde95,Avella11}).
\par
In the present paper we consider superconductivity in the extended Hubbard model  with a
weak intersite Coulomb repulsion $V_{ij}$ but in comparison with
Refs.~\cite{Raghu10,Alexandrov11,Raghu12}, we study  the limit of strong correlations, $U
\gg t$.  To compare various contributions to the superconducting $d$-wave pairing, we
consider also a model of the EPI  with strong forward scattering proposed in
Ref.~\cite{Zeyher96}.  The Dyson equation for the thermodynamic Green functions (GFs)
expressed in  terms  of the Hubbard operators (HOs) is derived using the Mori-type
projection technique~\cite{Mori65}. The self-energy is calculated in the noncrossing
approximation (NCA) as in the microscopic theory of the electronic spectrum in the normal
state in our previous publication~\cite{Plakida07}. We show that the kinematic
interaction for the HOs generates the AF superexchange pairing similar to the $t$--$J$
model. A contribution from the intersite Coulomb repulsion  in the first order suppresses
the pairing as found in Refs.~\cite{Alexandrov11,Raghu12}. But the kinematic interaction
induces also a strong electron interaction with spin-fluctuations which results  in the
$d$-wave superconductivity with high-$T_c$. Contribution from the EPI to the d-wave
pairing turned out to be small.
\par
In the next section we  introduce the model,  derive the Dyson equation, and calculate
the self-energy  in the NCA. A self-consistent system of equations is formulated  in
Sec.~\ref{sec:4}. Results of computations  of the electronic spectrum in the normal state
and of  superconducting $T_{\rm c}$ and the $d$-wave gap function are presented in
Sec.~\ref{sec:7}. Concluding remarks  are given in Sec.~\ref{sec:10}.  In the Appendix
details of the calculations are given.

\section{General formulation}
\label{sec:2}

\subsection{Extended Hubbard model}
\label{sec:2a}

We consider an extended Hubbard model   on a square lattice which we write  in terms of
the HOs~\cite{Hubbard65}:
\begin{eqnarray}
&& H= \varepsilon_1\sum_{i,\sigma}X_{i}^{\sigma \sigma} + \varepsilon_2\sum_{i}X_{i}^{22}
+  \sum_{i\neq j,\sigma}\bigl\{t_{ij}^{11}X_{i}^{\sigma 0}X_{j}^{0\sigma}
 \nonumber \\
&& + t_{ij}^{22}X_{i}^{2 \sigma}X_{j}^{\sigma 2} +  \sigma
t_{ij}^{12}(X_{i}^{2\bar\sigma}X_{j}^{0 \sigma} + {\rm H.c.})\bigr\} + H_{c, ep}.
 \label{2}
\end{eqnarray}
To apply the model  for consideration of the  cuprate superconductors, we introduce the
HOs for holes taking into account four  possible states on a lattice site $i$: an empty
state $(\alpha, \beta =0) $, a singly occupied hole state $(\alpha, \beta = \sigma)$ with
the spin
 $\sigma/2 = \pm (1/2)\,, \; \bar\sigma = -\sigma $,  and  a two-hole state  $(\alpha, \beta = 2) $. Then  the HO
$X_{i}^{\alpha\beta} = |i\alpha\rangle\langle i\beta|$ describes the transition from the
state $|i,\beta\rangle$ to the state $|i,\alpha\rangle$.  Energy  parameters in the model
(\ref{2}) are taken close to the values found  within the cell-perturbation
method~\cite{Feiner96} for the $p$-$d$ model for the CuO$_2$ plane~\cite{Emery87}.  In
particular, the single-particle energy $\varepsilon_1=\varepsilon_d-\mu$ is  the energy
of the $d$-type one-hole state measured from the chemical potential $\mu$  and the
two-particle energy $\varepsilon_2 = 2\varepsilon_1+ U $ is the energy of the two-hole
$p$-$d$ Zhang-Rice singlet state~\cite{Zhang88}. The effective Hubbard $U$ in cuprates is
the charge-transfer energy $ U = \Delta_{pd} = \epsilon_p-\epsilon_d$. According to the
cell perturbation method, in general case the values of the hopping parameters $t^{\eta
\zeta}_{ij}$ in (\ref{2}) depends  on the the subband indices $\eta, \zeta= 1, \,2$.
\par
In the last term in (\ref{2}) in addition to the inertsite CI $V_{ij}$ for holes in the
plane we take into account also  the EPI $g_{ij}$ for holes with phonons:
\begin{eqnarray}
H_{c, ep} &= & \frac{1}{2}  \sum_{i\neq j}\,V_{ij} N_i N_j + \sum_{i, j}\,g_{i j} N_i\,
u_j,
 \label{3}
\end{eqnarray}
where $u_j$ describes  an atomic displacement on the lattice site $j$ for a particular
phonon mode. More generally, the  EPI can be written as a sum $\sum_{\nu} g^{\nu}_{i,j}
u_{j}^{\nu}$ over the normal modes $\nu$. The hole number operator and the spin operators
in terms of HOs are defined as
\begin{eqnarray}
  N_i &=& \sum_{\sigma} X_{i}^{\sigma \sigma} + 2 X_{i}^{22},
\label{3a}\\
S_{i}^{\sigma} & = & X_{i}^{\sigma\bar\sigma} ,\quad
 S_{i}^{z} =  (\sigma/2) \,[ X_{i}^{\sigma \sigma}  -
  X_{i}^{\bar\sigma \bar\sigma}] .
\label{5}
\end{eqnarray}
The completeness relation for the HOs, $\, X_{i}^{00} +
 \sum_{\sigma} X_{i}^{\sigma\sigma}  + X_{i}^{22} = 1 $,
rigorously preserves the constraint of no double occupancy of the quantum state $\alpha$
on any lattice site $i$. From the multiplication rule $\, X_{i}^{\alpha\beta}
X_{i}^{\gamma\delta} = \delta_{\beta\gamma} X_{i}^{\alpha\delta} \,$ follows the
commutation relations:
\begin{equation}
\left[X_{i}^{\alpha\beta}, X_{j}^{\gamma\delta}\right]_{\pm}=
\delta_{ij}\left(\delta_{\beta\gamma}X_{i}^{\alpha\delta}\pm
\delta_{\delta\alpha}X_{i}^{\gamma\beta}\right)\, .
 \label{6}
\end{equation}
The upper sign  refers to the Fermi-type operators like $X_{i}^{0\sigma}$ while the lower
sign refers to the Bose-type operators like  $N_i$  (\ref{3a}) or the spin operators
(\ref{5}).
\par
The chemical potential $\mu$ depends on the average {\it hole} occupation number
\begin{equation}
  n =  1 + \delta = \langle \, N_i \rangle ,
    \label{4}
\end{equation}
where  $\langle ...\rangle$ denotes the statistical average with the Hamiltonian
(\ref{2}).
\par
We emphasize here that  the Hubbard model (\ref{2}) does not involve a dynamical coupling
of electrons (holes) to fluctuations of spins or charges. Its role is played by the {\it
kinematic} interaction caused by the complicated  commutation relations (\ref{6}), as was
already noted by Hubbard~\cite{Hubbard65}. For example, the equation of motion for the HO
$\, X\sb{i}\sp{\sigma 2}\, $ in the Heisenberg representation $(\hbar = 1)$ reads,
\begin{eqnarray}
 i\frac{d}{d t}  X\sb{i}\sp{\sigma 2} &= &[X\sb{i}\sp{\sigma 2}, H] =
   (\varepsilon_1 + U)\, X_{i}^{\sigma 2}\,
\nonumber \\
  &+& \sum\sb{l,\sigma '}\! \left( t\sb{il}\sp{22}
    B\sb{i\sigma\sigma '}\sp{22} X\sb{l}\sp{\sigma ' 2} -
    \sigma t\sp{21}\sb{il} B\sb{i\sigma\sigma '}\sp{21}
    X\sb{l}\sp{0\bar\sigma '} \right)
\nonumber \\
 &-& \sum\sb{l} X\sb{i}\sp{02} \left( t\sp{11}\sb{il}
    X\sb{l}\sp{\sigma0} +  \sigma t\sp{21}\sb{il}
    X\sb{l}\sp{2 \bar\sigma} \right)
\nonumber \\
& +&  \sum\sb{l}  X_{i}^{\sigma 2}( V_{i l}\,N_{l} +
 g_{il} \, u_{l} ).
\label{7}
\end{eqnarray}
Here $B\sb{i\sigma\sigma'}\sp{\eta \zeta}$  are the Bose-like operators,
\begin{eqnarray}
  B\sb{i\sigma\sigma'}\sp{22} & = & (X\sb{i}\sp{22} +
   X\sb{i}\sp{\sigma\sigma}) \, \delta\sb{\sigma'\sigma} +
   X\sb{i}\sp{\sigma\bar\sigma} \, \delta\sb{\sigma'\bar\sigma}
\label{8a} \\
  &  = &( N\sb{i}/2 +  \sigma\, S\sb{i}\sp{z}) \, \delta\sb{\sigma'\sigma} +
    S\sb{i}\sp{\sigma} \, \delta\sb{\sigma'\bar\sigma},
\nonumber \\
  B\sb{i\sigma\sigma'}\sp{21} & = & ( N\sb{i}/2 +
   \sigma S\sb{i}\sp{z}) \, \delta\sb{\sigma'\sigma} -
   S\sb{i}\sp{\sigma}\,  \delta\sb{\sigma'\bar\sigma},
  \label{8b}
\end{eqnarray}
where we used the definition of the number operator  (\ref{3a}) and the spin operators
(\ref{5}). The last term in (\ref{7}) is caused by the dynamic intersite CI and the  EPI.

\subsection{Dyson equation}
\label{sec:3}

To consider the superconducting pairing in the  model (\ref{2}), we introduce the
two-time thermodynamic  GF~\cite{Zubarev60} expressed in terms of the four-component
Nambu operators, $\, \hat X_{i\sigma}$ and   $\, \hat
X_{i\sigma}^{\dagger}=(X_{i}^{2\sigma}\,\, X_{i}^{\bar\sigma 0}\,\, X_{i}^{\bar\sigma
2}\,\, X_{i}^{0\sigma}) \,$:
\begin{eqnarray}
 {\sf G}_{ij\sigma}(t-t') & = & -i \theta(t-t')\langle \{
 \hat X_{i\sigma}(t) ,  \hat X_{j\sigma}^{\dagger}(t')\}\rangle
 \nonumber \\
 & \equiv & \langle \!\langle \hat X_{i\sigma}(t) \mid
    \hat X_{j\sigma}^{\dagger}(t')\rangle \!\rangle,
 \label{9}
\end{eqnarray}
where $ \{A, B\} = AB + BA$,  $ A(t)= \exp (i Ht) A\exp (-i Ht)$, and $\theta(x) = 1 $
for $x > 0 $ and $\theta(x) = 0 $ for $x < 0 $. The Fourier representation in $({\bf k},
\omega) $-space is defined by the relations:
\begin{eqnarray}
{\sf G}_{ij\sigma}(t-t') &= & \frac{1}{2\pi}\int_{-\infty}^{\infty} dt e^{- i(t-t')} {\sf
G}_{ij\sigma}(\omega),
 \label{9ft}\\
{\sf G}_{ij\sigma}(\omega) & = &\frac{1}{N}\,\sum_{\bf k}\exp[i{\bf k (i-j)}] {\sf
G}_{\sigma}({\bf k}, \omega).
    \label{9fk}
\end{eqnarray}
The GF (\ref{9ft}) is convenient to write in the matrix form
\begin{equation}
{\sf G}_{ij\sigma}(\omega)=
  {\hat G_{ij\sigma}(\omega)  \quad \quad
 \hat F_{ij\sigma}(\omega) \choose
 \hat F_{ij\sigma}^{\dagger}(\omega) \quad
   -\hat{G}_{ji\bar\sigma}(-\omega)} ,
 \label{9m}
\end{equation}
where the normal $\hat G_{ij\sigma}$ and  anomalous (pair) $\hat F_{ij\sigma}$  GFs are
the
 $2\times 2$ matrices for  two Hubbard subbands:
\begin{equation}
\hat G_{ij}(\omega) = \langle\! \langle \left(
\begin{array}{c}
     X_{i}^{\sigma2}  \\
     X_{i}^{0 \bar\sigma } \\
        \end{array}\right)  \mid
  X_{j}^{2\sigma} X_{j}^{\bar\sigma 0}
 \rangle \! \rangle_{\omega},
 \label{9a}
\end{equation}
\begin{equation}
\hat F_{ij}(\omega)  = \langle\! \langle \left(
\begin{array}{c}
     X_{i}^{\sigma2}  \\
     X_{i}^{0 \bar\sigma } \\
        \end{array}\right)  \mid
  X_{j}^{\bar\sigma 2} X_{j}^{0 \sigma}
 \rangle \! \rangle_{\omega}.
 \label{9b}
\end{equation}
\par
To calculate the GF (\ref{9})  we use the equation of motion method. Differentiating the
GF with respect to  time $t$, the Fourier representation of it leads to the equation
\begin{equation}
 \omega {\sf G}\sb{ij\sigma}(\omega) = \delta \sb{ij} {\sf Q} +
   \langle \!\langle
    [\hat X\sb{i\sigma},H] \mid  \hat X\sb{j\sigma}\sp{\dagger}
   \rangle \!\rangle\sb{\omega}\, .
\label{10}
\end{equation}
Here the  correlation function ${\sf Q} = \langle \{\hat X\sb{i\sigma},\hat
X\sb{i\sigma}\sp{\dagger}\}\rangle   = \hat{\tau}_{0} \times \hat Q\,$  where   $\, \hat
Q =
\left(  \begin{array}{cc} Q\sb{2} & 0 \\
      0 & Q\sb{1} \end{array}  \right)\,$
and $\hat{\tau}_{0}$ is the $2 \times 2$ unit matrix. The spectral weights of the Hubbard
subbands  in the paramagnetic state  $\, Q\sb{2} = \langle X\sb{i}\sp{22} +
X\sb{i}\sp{\sigma\sigma} \rangle = n/2 \,$ and $\, Q\sb{1} = \langle X\sb{i}\sp{00} +
X\sb{i}\sp{\bar\sigma \bar\sigma} \rangle = 1-Q\sb{2}\, $ depend on the occupation number
of holes (\ref{4}). In  the $\, {\sf Q}$ matrix we neglect anomalous averages of the type
$\langle X_i^{02} \rangle$ which are irrelevant for  the $d$-wave pairing~\cite{Adam07}.
\par
To introduce the zero-order quasiparticle (QP) excitation energy we use the Mori-type
projection method~\cite{Mori65}. In this approach, the many-particle operator $\hat
Z\sb{i\sigma} = [\hat X\sb{i\sigma},H]$ in (\ref{10}) is written as  a sum of a linear
part and an irreducible part $\hat Z\sb{i\sigma}\sp{(ir)}$ orthogonal to $\hat
X\sb{j\sigma}\sp{\dagger}$:
\begin{equation}
  \hat Z\sb{i\sigma} = [\hat X\sb{i\sigma}, H] =
    \sum\sb{l}{\sf E}\sb{il\sigma} \hat X\sb{l\sigma} +
    \hat Z\sb{i\sigma}\sp{(\rm ir)}.
\label{11}
\end{equation}
The orthogonality condition
 $\langle \{ \hat Z\sb{i\sigma}\sp{(\rm ir)}, \,
 \hat X\sb{j\sigma}\sp{\dagger}\} \rangle =  0 \, $
determines the excitation  energy in the mean-field approximation (MFA)
\begin{eqnarray}
 {\sf E}\sb{ij\sigma} &=&  \langle \{ [\hat X\sb{i\sigma}, H],
    \hat X\sb{j\sigma}\sp{\dagger} \} \rangle {\sf Q}^{-1}
 \nonumber\\
   &= & (1/N)\,\sum_{\bf k}\exp[i{\bf k (i-j)}]\,
   {\sf E}\sb{\sigma}({\bf k}),
\label{12}
\end{eqnarray}
and the corresponding zero-order GF
\begin{equation}
  {\sf G}\sp{0}\sb{\sigma }({\bf k},\omega) =
    \Bigl( \omega \tilde{\tau}\sb{0} - {\sf E}\sb{\sigma}({\bf k})
      \Bigr) \sp{-1} {\sf Q} \, ,
\label{13}
\end{equation}
where $\tilde{\tau}\sb{0}$ is the $4\times 4$ unit matrix.
\par
To calculate the multiparticle GF $\,\langle \langle
    \hat Z\sb{i\sigma}\sp{(\rm ir)} (t) \mid  \hat X\sb{j\sigma}\sp{\dagger}(t')
   \rangle \rangle \,$ in (\ref{10}) we differentiate it with respect
to the second time $t'$ and apply the same projection procedure as in (\ref{11}). This
results in the equation for the GF (\ref{10}) in  the form,
\begin{equation}
 {\sf G}\sb{\sigma}({\bf k}, \omega) =
 {\sf G}\sp{0}\sb{\sigma }({\bf k},\omega) + {\sf G}\sp{0}\sb{\sigma }({\bf
 k},\omega)\,{\sf T}\sb{\sigma }({\bf k},\omega)\,
 {\sf G}\sp{0}\sb{\sigma }({\bf k},\omega),
\label{14a}
\end{equation}
where the scattering matrix
\begin{equation}
  {\sf T}\sb{\sigma}({\bf k}, \omega) =
 {\sf  Q}\sp{-1}\,\langle\!\langle {\hat Z}\sb{{\bf k}\sigma}\sp{(\rm ir)} \!\mid\!
     {\hat Z}\sb{{\bf k}\sigma}\sp{(\rm ir)\dagger} \rangle\!\rangle
      \sb{\omega}\;{\sf  Q}\sp{-1} .
\label{15a}
\end{equation}
Now we can introduce the self-energy operator ${\sf \Sigma}\sb{\sigma}({\bf q}, \omega)$
as the {\it proper} part ({\rm pp}) of the scattering matrix (\ref{15a}) which has no
parts connected by the zero-order GF (\ref{13}) according to the equation: ${\sf T} =
{\sf \Sigma} + {\sf \Sigma } \, {\sf G}\sp{0} \, {\sf T}$. The definition of the proper
part of the scattering matrix (\ref{15a}) is equivalent to an introduction of a projected
Liouvillian superoperator for the memory function in the conventional Mori
technique~\cite{Mori65}.
\par
Using the self-energy operator instead of the scattering matrix in Eq.~(\ref{14a}) we
obtain the Dyson equation for the GF (\ref{9}):
\begin{equation}
 {\sf G}\sb{\sigma}({\bf k}, \omega) =
  \left[\omega \tilde{\tau}\sb{0} - {\sf E}\sb{\sigma}({\bf k})
  -    {\sf  Q} {\sf \Sigma}_{\sigma}({\bf k}, \omega)
  \right] \sp{-1} {\sf Q},
\label{14}
\end{equation}
where the self-energy operator is given by
\begin{equation}
 {\sf  Q} {\sf \Sigma}\sb{\sigma}({\bf k}, \omega) =
    \langle\!\langle {\hat Z}\sb{{\bf k}\sigma}\sp{(\rm ir)} \!\mid\!
     {\hat Z}\sb{{\bf k}\sigma}\sp{(\rm ir)\dagger} \rangle\!\rangle
      \sp{(\rm pp)}\sb{\omega}\;{\sf  Q}\sp{-1} .
\label{15}
\end{equation}
Dyson equation  (\ref{14})  with the zero-order QP excitation energy (\ref{12}) and the
self-energy (\ref{15}) gives an exact representation for the GF (\ref{9}).  The
self-energy takes into account processes of inelastic scattering of electrons (holes) on
spin  and charge fluctuations due to the kinematic interaction and the dynamic intersite
CI and the EPI (see Eq.~(\ref{7})). To obtain a closed system of equations, the
multiparticle GF in the self-energy operator (\ref{15}) should be evaluated as discussed
below.

\section{Self-consistent system of equations}
\label{sec:4}

\subsection{Mean-field approximation}
\label{sec:4a}

The superconducting pairing in the Hubbard model already occurs in  the MFA  and is
caused by the  kinetic superexchange interaction as in the $t$--$J$
model~\cite{Anderson87}. Therefore, it is reasonable to consider at first the MFA
described by the zero-order GF (\ref{13}). Using commutation relations (\ref{6}) for the
HOs  we calculate the energy matrix (\ref{12}):
\begin{equation}
{\sf E}\sb{ij\sigma} = \left(
\begin{array}{cc}
  \hat{\varepsilon}_{ij}   & \hat{\Delta}_{ij\sigma} \\
     \hat{\Delta}_{ji\sigma}^{*} &
     -\hat{\varepsilon}_{ji\bar\sigma}
\end{array}\right).
 \label{16}
\end{equation}
The matrix $\hat{\varepsilon}({\bf k}) = \sum_j\,\exp[i{\bf k} ({\bf i-j})]\,
 \hat{\varepsilon}_{ij}$ after diagonalization determines the QP spectrum in the two
Hubbard subbands in the normal state (for details see~\cite{Plakida07}):
\begin{eqnarray}
{\varepsilon}_{1, 2} ({\bf k})& = & ({1}/{2}) [\omega_{2} ({\bf k}) + \omega_1 ({\bf k})]
\mp({1}/{2}) \Lambda({\bf k}),
 \label{17}\\
 {\omega}_\iota({\bf k})& = &
 4  t\,\alpha_{\iota} \gamma({\bf k})
 + 4 \,\beta_{\iota}\,t'\gamma'({\bf k})+
 4 \,\beta_{\iota}\,t''\gamma''({\bf k})
 \nonumber   \\
& + &\omega^{(c)}_\iota({\bf k}) + U \delta_{\iota,2} - \mu,
 \quad (\iota = 1, 2)
\label{17a}\\
  \Lambda({\bf k}) &= &    \{[\omega_{2} ({\bf k})
  - \omega_1 ({\bf k})]^2 + 4 W({\bf k})^2 \}^{1/2},
\nonumber\\
 W({\bf k}) & = &  4  t\,\alpha_{12} \gamma({\bf k})
 + 4 t' \,\beta_{12} \gamma'({\bf k})+
 4 t'' \,\beta_{12} \gamma''({\bf k}).
\nonumber
\end{eqnarray}
Here  the hopping parameters  in (\ref{2}) are assumed  to be equal, $\, t^{22}_{ij} =
t^{11}_{ij} = t^{12}_{ij} = t_{ij}$, where $t_{ij}$ is defined by the expression:
\begin{eqnarray}
t_{ij} & = & (1/N)\,\sum_{\bf k}\exp[i{\bf k (i-j)}]\, t({\bf k}),
\label{17b}\\
t({\bf k})& =&  4 t \, \gamma({\bf k}) + 4 t' \,\gamma'({\bf k}) + 4 t''\, \gamma''({\bf
k}).
 \label{17c}
\end{eqnarray}
The hopping parameters are equal to  $t\,$ for the nearest neighbor (n.n.) sites $ a_{1}
= ( \pm a_{x}, \pm a_{y})$,  $\, t'\,$ -- for the next nearest neighbor (n.n.n.) sites
$a_d = \pm (a_x \pm a_y)$, and $ \, t''\,$ -- for the  n.n.n. sites $\,a_{2}= \pm 2
a_{x}, \pm 2 a_{y}$. The corresponding  ${\bf k}$-dependent functions are: $\,\gamma({\bf
k})= (1/2)(\cos k_x +\cos k_y), \; \gamma'({\bf k}) = \,\cos k_x \cos k_y \, $, and $\;
\gamma '' ({\bf k})= (1/2)(\cos 2 k_x +\cos 2 k_y) $ (the lattice constants $ a_{x}=
a_{y}$ are put to unity). The contribution from the  CI $V_{i j}$ in (\ref{17a})  is
given by
\begin{equation}
\omega^{(c)}_{1(2)}({\bf k})= \frac{1}{N } \sum_{\bf q} V({\bf k -q}) N_{1(2)}({\bf q}),
 \label{17ci}
 \end{equation}
where $ N_1({\bf q}) = \langle X_{\bf q}^{0 \bar\sigma}X_{\bf q}^{\bar\sigma 0}\rangle /
Q_1, \;  N_2({\bf q})=
 \langle X_{\bf q}^{\sigma 2}X_{_{\bf q}}^{2\sigma}\rangle / Q_2$
and $V({\bf k -q})$ is the Fourier transform of $V_{ij}$.
\par
The kinematic interaction for the HOs results in renormalization of  the spectrum
(\ref{17}) determined by the parameters: $\, \alpha_{\iota}= Q_{\iota}[ 1 +
{C_{1}}/{Q^2_{\iota}}], \, \beta_{\iota} = Q_{\iota}[ 1 + {C_{2}}/{Q^2_{\iota}}]\,$, $\,
\alpha_{12}= \sqrt{Q_{1}Q_{2}}[ 1 - {C_{1}}/{Q_{1}Q_{2}}] ,\, \beta_{12} =
\sqrt{Q_{1}Q_{2}}[ 1 -{C_{2}}/{Q_{1}Q_{2}}]\,$. Here we  take into account the
renormalization caused by the spin correlation functions for the n.n. and the n.n.n.
sites, respectively:
\begin{equation}
C_{1} = \langle {\bf S}_i{\bf S}_{i+ a_1} \rangle, \quad C_{2} = \langle {\bf S}_i{\bf
S}_{i+ a_d}\rangle \approx  \langle {\bf S}_i{\bf S}_{i+ a_2}\rangle  .
 \label{18}
\end{equation}
The short-range AF correlations  considerably suppress the n.n. hopping parameters since
$C_1 < 0\,$ and at low doping $\,|C_1| = 0.1 - 0.2$  that results in $\alpha_{\iota}\ll
1$. At the same time,  the n.n.n. hopping parameters are increased since $C_2 > 0$.
\par
Now we evaluate the anomalous components $\hat{\Delta}_{ij\sigma}$ of the matrix
(\ref{16}) which determine the superconducting gap in the MFA. Considering  the singlet
$d$-wave pairing, we calculate the intersite pair correlation functions. The diagonal
matrix components are given by the equations:
\begin{eqnarray}
&&   \Delta\sb{ij\sigma}\sp{22} Q_2 =  -  \sigma\, t_{ij}^{21}
   \langle X\sb{i}\sp{02} N\sb{j} \rangle  - V_{i j}
 \langle  X_{i}^{\sigma 2}\, X_{j}^{\bar\sigma 2} \rangle ,
   \label{19a}\\
&&  \Delta\sb{ij\sigma}\sp{11} Q_1 =   \sigma \, t_{ij}^{12}
   \langle N\sb{j}  X\sb{i}\sp{02} \rangle - V_{i j}
   \langle  X_{i}^{0 \bar\sigma} X_{j}^{0\sigma}  \rangle .
 \label{19b}
\end{eqnarray}
Here we used the original notation for the interband hopping parameters $ t_{ij}^{12}$ to
emphasize that the kinematic pairing $\langle X\sb{i}\sp{02} N\sb{j} \rangle$ is mediated
by the interband hopping.  In terms of  the Fermi operators  $ a_{i\sigma} =
X_{i}^{0\sigma} + \sigma X_{i}^{\bar\sigma 2}$, the pair correlation function  in
(\ref{19a}) can be written as $\, \langle X\sb{i}\sp{02} N\sb{j} \rangle  = \langle
X\sb{i}\sp{0\downarrow} X\sb{i}\sp{\downarrow 2} N\sb{j}\rangle =\langle
a\sb{i\downarrow}\, a\sb{i\uparrow} N\sb{j} \rangle $. This representation  shows that
the kinematic pairing occurs on a single lattice site but in two Hubbard
subbands~\cite{Plakida03}.
\par
The  correlation function $\, \langle X\sb{i}\sp{02} N\sb{j} \rangle$ can be calculated
directly from the    GF $\, L_{ij}(t-t') = \langle \langle X_{i}^{02} (t) \mid N_j (t')
\rangle \rangle \, $ without {\it any decoupling} approximation as shown in
Ref.~\cite{Plakida03}. In particular, under hole doping, $ n = 1 + \delta > 1$, the pair
correlation function in the two-site approximation reads (for details see Appendix A):
\begin{eqnarray}
  \langle X_{i}^{02} N_j \rangle =
   -\frac{4  t_{ij}^{12} }{U}  \sigma \,
  \langle X_{i}^{\sigma2}X_{j}^{\bar\sigma2} \rangle .
 \label{20}
\end{eqnarray}
As a result, the equation for the superconducting gap in (\ref{19a})  can be written as
\begin{eqnarray}
\Delta^{22}_{ij\sigma}  =
 (J_{i j} - V_{i j})\,\langle X_{i}^{\sigma2}
 X_{j}^{\bar\sigma2}\rangle /Q_2 ,
 \label{21}
\end{eqnarray}
where $\, J_{i j} = {4\, ( t_{ij}^{12})^2}/{U }$ is the AF superexchange interaction.  A
similar equation holds  for the gap in the one-hole subband:
 $\, \Delta_{ij\sigma}^{11}= (J_{i j} -
 V_{i j})\,\langle X_{i}^{0\bar\sigma}
 X_{j}^{0\sigma} \rangle/Q_1 \,$. We thus conclude that the
pairing in the Hubbard model in the MFA  is similar to  the superconductivity in the
$t$--$J$ model mediated by the  AF superexchange interaction $J_{i j}$.

\subsection{Self-energy operator}
\label{sec:5}

Self-energy operator (\ref{15}) can be  written in the same matrix form as the  GF
(\ref{9m}):
\begin{equation}
{\sf Q}\, {\sf \Sigma}_{ij\sigma}(\omega) =  {\hat M_{ij\sigma}(\omega) \quad  \quad
\hat\Phi_{ij\sigma}(\omega) \choose \hat\Phi_{ij\sigma}^{\dagger} (\omega)\quad
-\hat{M}_{ji\bar\sigma}(-\omega)} {\sf Q}^{-1}  \, ,
 \label{23}
\end{equation}
where the matrices $\hat M$ and $\hat\Phi$  denote the respective normal and anomalous
(pair) components of the self-energy operator. The system of equations for the $(4 \times
4)$ matrix GF (\ref{9m}) and the self-energy (\ref{23}) can be reduced to   a system of
equations for the normal ${\hat G}_\sigma({\bf k},\omega)$ and the pair ${\hat
F}_\sigma({\bf k},\omega)$ $(2 \times 2)$ matrix components. Using representations for
the energy matrix (\ref{16}) and the self-energy (\ref{23}), we derive  for these
components  the following system of matrix equations:
\begin{eqnarray}
{\hat G}({\bf k},\omega) & = & \Bigl(
  \hat {G}_{N}({\bf k},\omega)^{-1}
\nonumber \\
& +  & \hat{\varphi}_\sigma({\bf k},\omega)\,
  \hat{G}_{N}({\bf k},- \omega)\,\hat{\varphi}^{*}_\sigma({\bf
k},\omega)  \Bigr)^{-1} \, \hat{Q}, \qquad
 \label{24} \\
{\hat F}_\sigma({\bf k},\omega) & = & -\hat{G}_{N}({\bf k},-
\omega)\,\hat{\varphi}_\sigma({\bf k},\omega) \,
 \hat{G}({\bf k},\omega) ,
 \label{25}
\end{eqnarray}
where we introduced  the normal state  GF
\begin{eqnarray}
{\hat G}_{N}({\bf k},\omega)& = & \Bigl( \omega \hat\tau_0 - \hat{\varepsilon}({\bf k}) -
  \hat{M}({\bf k},\omega)/ \hat{Q} \Bigr)^{-1},
\label{26}
 \end{eqnarray}
and the  superconducting gap function
\begin{eqnarray}
{\hat \varphi}_\sigma({\bf k},\omega)& = & \hat{\Delta}_{\sigma}({\bf k}) +
 \hat\Phi_{\sigma}({\bf k},\omega) /\hat{Q} .
 \label{27}
\end{eqnarray}
To calculate  the self-energy matrix~(\ref{23}) we use the mode-coupling approximation
which is equivalent to the NCA in the diagram technique for GFs. In this approximation, a
propagation of Fermi-like excitations described by the operator $\, X_l^{\sigma' 2}, \,$
and Bose-like excitations described by the operator $B\sb{i\sigma\sigma'}$ for $l \neq i$
is assumed  to be independent. Therefore, the time-dependent multiparticle correlation
functions in the self-energy operators (\ref{23}) can be written as a product of
fermionic and bosonic correlation functions,
\begin{eqnarray}
 &&\langle X_{l'}^{2\sigma''} B\sb{j\sigma\sigma''}^\dag
 |B\sb{i\sigma\sigma'}(t)
X_l^{\sigma' 2}(t)\rangle
\nonumber \\
 && = \delta_{\sigma', \sigma''}\langle X_{l'}^{2\sigma'}
X_l^{\sigma' 2}(t)\rangle \langle B\sb{j\sigma\sigma'}^\dag
 |B\sb{i\sigma\sigma'}(t) \rangle,
\label{B5} \\
&& \langle  X_{l'}^{\bar{\sigma}'' 2} B\sb{j \bar{\sigma}\bar{\sigma}''}
 | B\sb{i\sigma \sigma'}(t) X_l^{\sigma' 2}(t)\rangle
\nonumber \\
 && = \delta_{\sigma', \sigma''}
 \langle  X_{l'}^{\bar{\sigma}' 2}
  X_l^{\sigma' 2}(t)\rangle\,
  \langle   B\sb{j\bar{\sigma}\bar{\sigma}'}
  B\sb{i\sigma \sigma'}(t) \rangle \, .
 \label{B6}
\end{eqnarray}
The time-dependent correlation functions are calculated self-consistently using the
corresponding GFs (for details see Appendix B).
\par
In particular, the normal and anomalous diagonal components of the self-energy for the
two-hole subband are determined by the expressions
\begin{eqnarray}
 M\sp{22}({\bf k},\omega) &= &
   \frac{1}{N} \sum\sb{\bf q}
   \int\limits\sb{-\infty}\sp{+\infty} \!\!{\rm d}z\,
   K^{(+)}(\omega,z|{\bf q },{\bf k-q})
   \nonumber \\
& \times &\left\{ - \frac{1}{\pi} \mbox{Im} \left[
     G\sp{22}({\bf q},z) +
   G\sp{11}({\bf q},z) \right] \right\} , \quad
 \label{29}
\end{eqnarray}
\begin{eqnarray}
 \Phi\sb{\sigma}\sp{22}({\bf k},\omega) & = &
 \frac{1}{N} \sum\sb{\bf q}
   \int\limits\sb{-\infty}\sp{+\infty} \!\!{\rm d}z\,
   K^{(-)}(\omega,z|{\bf q },{\bf k-q})
  \nonumber \\
& \times & \left\{ - \frac{1}{\pi} \mbox{Im} \left[
   F\sb{\sigma}\sp{22}({\bf q},z) -
 F\sb{\sigma }\sp{11}({\bf q},z)
 \right]\right\} ,\quad
 \label{30}
\end{eqnarray}
where $ G\sp{\alpha\alpha}({\bf q},z)$ and $F\sb{\sigma }\sp{\alpha\alpha}({\bf q},z) $
are given by the diagonal components of the matrices (\ref{24}), (\ref{25}). Similar
expressions hold for the  self-energy components $M\sp{11}({\bf k},\omega)$ and
$\Phi\sb{\sigma}\sp{11}({\bf k},\omega)$ for electron doping when the  Fermi energy
located in the one-hole subband (see Ref.~\cite{Plakida97}). Note, that in the
paramagnetic normal state the GF (\ref{24}) and the self-energy (\ref{29}) do not depend
on the spin $\sigma$.
\par
The kernel of the integral equations (\ref{29}), (\ref{30}) has a form, similar to the
strong-coupling Migdal-Eliashberg theory~\cite{Migdal58,Eliashberg60}:
\begin{eqnarray}
&&  K^{(\pm)}(\omega,z |{\bf q },{\bf k -q })  =
\frac{1}{\pi}\int\limits\sb{-\infty}\sp{+\infty} {\rm d}\Omega
  \, \frac{1 +N(\Omega) - n(z)}{\omega - z - \Omega}
\nonumber \\
&&\!\!\!\!\!\!  \times \big\{ |t({\bf q})|^{2} {\rm Im} \chi\sb{sf}({\bf k- q},\Omega)
\pm |g({\bf k -q})|^2 {\rm Im} \chi_{ph}({\bf k-q}, \Omega)
\nonumber \\
&& \pm \left[ |V({\bf k -q})|^2 + | t({\bf q})|^{2}/4 \right]
 {\rm Im}\, \chi_{cf}({\bf k-q}, \Omega)\big\},
  \label{31}
\end{eqnarray}
where $\,n(\omega)= [e^{\omega/T} +1]^{-1}$ and $\,N(\omega)= [e^{\omega/T} -1]^{-1}$.
The  spectral densities of bosonic excitations are determined by the dynamic
susceptibility for spin $( sf)$, number (charge) $( cf)$, and lattice (phonon) $( ph)$
fluctuations
\begin{eqnarray}
\chi\sb{sf}({\bf q},\omega) & = &
  -  \langle\!\langle {\bf S\sb{q} | S\sb{-q}}
\rangle\!\rangle\sb{\omega},
\label{32a} \\
 \chi\sb{cf}({\bf q},\omega) &= &
 - \langle\!\langle \delta N\sb{\bf q} | \delta N\sb{-\bf q}
   \rangle\!\rangle\sb{\omega} ,
\label{32b}\\
\chi\sb{ph}({\bf q},\omega) &= &
 - \langle\!\langle  u\sb{\bf q} | u\sb{-\bf q}
   \rangle\!\rangle\sb{\omega} ,
\label{32c}
\end{eqnarray}
which are defined in terms of the commutator GFs~\cite{Zubarev60} for the spin ${\bf S
\sb{q}} $,  number $\delta N_{\bf q} = N_{\bf q} - \langle N_{\bf q} \rangle$, and
lattice displacement (phonon) $ u\sb{\bf q} $  operators.
\par
In the NCA,  vertex corrections  are neglected as in the Migdal-Eliashberg theory. For
the EPI $\, g({\bf k -q})\,$  the vertex corrections are small, as shown by
Migdal~\cite{Migdal58}. The  interaction $\, t({\bf q}) \,$ with spin-fluctuations
(\ref{32a}) induced by the intraband hopping is not small and vertex corrections may be
important. However, as was shown in Ref.~\cite{Liu92}  a certain set of diagrams, in
particular the first crossing diagram, vanishes due to kinematic restrictions for spin
scattering processes. Moreover, in Ref.~\cite{Monthoux97} it was found that the
renormalization of the vertex for a short AF correlation length is weak.  Therefore, the
NCA for the self-energy calculated self-consistently  can be considered as a reasonable
approximation. This approximation makes it possible to consider the strong coupling
regime  which is essential in study of  renormalization of quasiparticle spectra and  the
superconducting pairing.

\subsection{Two-subband model}
\label{sec:6}

In this section we derive  a self-consistent system of equations for the GFs
(\ref{24})--(\ref{26}) and the self-energy components (\ref{29}), (\ref{30})  for the two
Hubbard subbands adopting several approximations to make the system of equations
numerically tractable.
\par
At first we consider equations for the normal state. The diagonal components of the GF
(\ref{26}) can be written as~\cite{Plakida07}:
\begin{eqnarray}
 {G}^{11(22)}_{N}({\bf k},\omega)& =&
[1 - b({\bf k})] {G}_{1(2)}({\bf k},\omega) \nonumber \\
 &+&   b({\bf k}){G}_{2(1)}({\bf k},\omega)  ,
 \label{33}\\
{G}_{1 (2)}({\bf k},\omega)& = & \frac{1}
 {\omega - {\varepsilon}_{1(2)}({\bf k}) -
 \Sigma({\bf k},\omega)} \, ,
 \label{34}
\end{eqnarray}
where the hybridization parameter $\,  b({\bf k}) = [{\varepsilon}_{2} ({\bf k}) -
\omega_{2}({\bf k})] / [{\varepsilon}_{2} ({\bf k}) - {\varepsilon}_{1} ({\bf k})]\, $.
The self-energy can be approximated by the same  function for the both subbands,
\begin{eqnarray}
 \Sigma({\bf k},\omega) &= &
 \frac{1}{N} \sum\sb{\bf q}
   \int\limits\sb{-\infty}\sp{+\infty} \!\!{\rm d}z\,
   K^{(+)}(\omega,z|{\bf q },{\bf k-q})
   \nonumber \\
& \times & [ - ({1}/{\pi})]\, \mbox{Im}\,
 [ G_{1}({\bf q},z) + G_{2}({\bf q},z) ] .
 \label{35}
\end{eqnarray}
The chemical potential $\mu$  is calculated from the equation for the average  hole
occupation number (\ref{4}):
\begin{equation}
  n =  1 + \delta = 2 \langle X_{i}^{\sigma \sigma}\rangle
  + 2 \langle X_{i}^{22}\rangle
    = \frac{2}{N} \sum_{{\bf q}} N_{h}({\bf q}) ,
\label{36}
\end{equation}
where the hole occupation number is given by
\begin{eqnarray}
N_{(h)}({\bf k}) &=& N_{(h 1)}({\bf k}) + N_{(h 2)}({\bf k}),
 \label{37} \\
N_{(h 1)}({\bf k}) &=& [Q_1 + (n-1)b({\bf k})]\, {N}_{1}({\bf k}),
\nonumber \\
N_{(h 2)}({\bf k})  &=& [ Q_2 - (n-1)b({\bf k})]\, {N}_{2}({\bf k}),
\nonumber \\
{N}_{1(2)}({\bf k})&=& -\frac{1}{\pi}\,\int^{\infty}_{-\infty}
\frac{d\omega}{e^{\omega/T} +1}\, \mbox{Im}\, { G}_{1(2)}({\bf k},\omega)\, .
 \label{38}
\end{eqnarray}
Density of  states (DOS) is determined by
\begin{equation}
A(\omega) = \frac{1}{N } \sum_{\bf k}\, {A}_{(h)}({\bf k}, \omega) ,
 \label{39}
\end{equation}
where the spectral function for holes reads
\begin{eqnarray}
 {A}_{(h)}({\bf k}, \omega) & = &  [Q_1 +  P({\bf k})]
 {A}_{1}({\bf k}, \omega)
 \nonumber \\
 &+&[Q_2 -  P({\bf k})]{A}_{2}({\bf k}, \omega),
\label{40} \\
A_{1 (2)}({\bf k}, \omega) &=&
 - ({1}/{\pi})\, {\rm Im}{G}_{1 (2)}({\bf k},\omega).
\nonumber
\end{eqnarray}
Here the hybridization parameter $\,P({\bf k}) =(n-1) b({\bf k}) - 2 \sqrt{Q_1\, Q_2}$ $[
W({\bf k})/\Lambda({\bf k})]\,$ takes into account contributions  from both the diagonal
and off-diagonal components of the GF  (\ref{26}).
\par
Now we formulate equations for the superconducting gap (\ref{27}). We consider the case
of hole doping  when the Fermi energy is located in the two-hole subband, $n = 1 +\delta
> 1$. By taking into account the gap equation (\ref{21}) in the MFA
and the self-energy (\ref{30}), Eq.~(\ref{27}) for the two-hole subband gap $\varphi({\bf
k},\omega)= \sigma \varphi_{2, \sigma}({\bf k},\omega)$ reads,
\begin{eqnarray}
\varphi({\bf k},\omega) & = &
 \frac{1}{N} \sum_{\bf q}
  \int\limits\sb{-\infty}\sp{+\infty}\, d z \,[- \frac{\sigma}{\pi Q_2}
{\rm Im} F^{22}\sb{\sigma}({\bf q},z)]
 \nonumber \\
&\times& \big\{[ J({\bf k-q}) - V({\bf k-q})]\, \frac{1}{2}\tanh\frac{z}{2T}
\nonumber \\
 & + & K^{(-)}(\omega,z|{\bf q },{\bf k-q})\big\}.
      \label{42}
\end{eqnarray}
Here the contribution from the one-hole subband $F^{11}_{\sigma}({\bf q},z)$ in
(\ref{30}) was neglected since this filled band much below the Fermi level gives a
vanishingly small contribution  to the pairing.  To determine the superconducting $T_{\rm
c}$ it is sufficient to solve a linear equation for the gap (\ref{42}) using the linear
approximation for the pair  GF (\ref{25}),
\begin{eqnarray}
&& F^{22}_\sigma({\bf k},\omega) =  - G_{N}^{22}({\bf k},-\omega)
 G_{N}^{22}({\bf k},\omega)\,\sigma {\varphi}({\bf k},\omega)
 Q_{2}
\nonumber \\
&& \approx - [1 - b({\bf q})]^2\,G_{2}({\bf k},-\omega) G_{2}({\bf k},\omega)\, \sigma
{\varphi}({\bf k},\omega)  Q_{2}. \quad
 \label{43}
\end{eqnarray}
As in (\ref{42}), here we neglect  the GF ${G}_{1}({\bf k},\omega)$ in  (\ref{33}) since
the  contribution  to the pairing from the one-hole subband much below the Fermi energy
is small.
\par
For numerical calculations, it is convenient to introduce the imaginary frequency
representation, $ i\omega_{n}=i\pi T(2n+1)$, $n = 0,\pm 1, \pm 2, ...\, $.  In this
representation the self-energy (\ref{35}) reads
\begin{eqnarray}
{\Sigma}({\bf k}, \omega_{n}) & = & - \frac{T}{N}\sum_{\bf q}
 \sum_{m}\lambda^{(+)}({\bf q, k-q} \mid
\omega_{n}-\omega_{m})
 \nonumber \\
&\times &    [{G}_{1}({\bf q}, \omega_{m})+
  {G}_2({\bf q}, \omega_{m})]
\nonumber\\
 &\equiv & i\omega_{n}\,[1-Z({\bf k},\omega_n)]
  + X({\bf k},\omega_n).
 \label{45a}
\end{eqnarray}
Here we introduced the  renormalization parameters
\begin{eqnarray}
\!\! \!\! \omega\; [1- Z({\bf k},\omega)]= (1/2)\,
 [{\Sigma}({\bf k}, \omega)
   - {\Sigma}({\bf k}, -\omega)]  , \; &&
\label{45b}\\
 X({\bf k},\omega)
=(1/2)\,[{\Sigma}({\bf k}, \omega)
   + {\Sigma}({\bf k}, -\omega)]  . \;&&
\label{45c}
 \end{eqnarray}
The normal GF (\ref{34}) for the two subbands  takes the form:
\begin{eqnarray}
\{G_{1(2)}({\bf k},  \omega_{n})\}^{-1}= i\omega_n  - {\varepsilon}_{1(2)}({\bf k}) -
\Sigma({\bf k},\omega_n)
 \nonumber \\
 = i\omega_n Z({\bf k},\omega_n) -[{\varepsilon}_{1(2)}({\bf k})+
X({\bf k},\omega_n)] \, .
 \label{45}
\end{eqnarray}
The hole occupation number (\ref{38})  in terms of  the GF (\ref{45}) reads:
\begin{eqnarray}
{N}_{1(2)}({\bf k})= \frac{1}{2}+ T\sum_{m} \,
   G_{1(2)}({\bf k}, \omega_{m}).
 \label{38a}
\end{eqnarray}
The   gap equation (\ref{42})  in the linear approximation for the pair GF (\ref{43}) can
be written as
\begin{eqnarray}
\varphi({\bf k}, \omega_n) & = &
  \frac{T_c}{N}\sum_{\bf q} \,  \sum_{m}\,
\{\, J({\bf k-q}) - V({\bf k-q})
\nonumber \\
& +  & \lambda^{(-)}({\bf q, k-q} \mid \omega_{n}-\omega_{m}) \}
 \label{46} \\
 &\times & \frac{[1 - b({\bf q})]^2\,
  \varphi({\bf q}, \omega_{m})}{[\omega_m Z({\bf q},\omega_m)]^2
  + [{\varepsilon}_{2}({\bf q})+
X{\bf q},\omega_m)]^2}\; .
 \nonumber
 \end{eqnarray}
The  interaction functions in (\ref{45a}) and (\ref{46}) in the imaginary frequency
representation are given by
\begin{eqnarray}
 && \lambda^{(\pm)}({\bf q },{\bf k -q } | \nu_n)  = -
|t({\bf q})|^{2} \, \chi\sb{sf}({\bf k- q},\nu_n)
\nonumber \\
&& \mp \big\{\left[ |V({\bf k -q})|^2 + | t({\bf q})|^{2}/4 \right]\chi_{cf}({\bf k-q},
\nu_n)
\nonumber \\
&& + |g({\bf k -q})|^2 \, \chi_{ph}({\bf k-q}, \nu_n)
  \,  \big\}\, .
  \label{47}
\end{eqnarray}
Thus, we have derived  the self-consistent system of equations for the normal GF
(\ref{45}), the self-energy (\ref{45a}), and the gap function (\ref{46}).

\section{Results and discussion}
\label{sec:7}
\subsection{Model parameters}
\label{sec:7a}

To perform numerical calculations we should specify model parameters in the derived
system of equations. For the intersite CI $\,V({\bf q}) \,$ we consider two models. In
the first model the CI is determined by the repulsion of two holes on the n.n. lattice
sites,
\begin{equation}
V_1({\bf q}) = 2 V_1\, (\cos q_x + \cos q_y ) .
 \label{50aa}
\end{equation}
According to the cell-perturbation method~\cite{Feiner96}, for the conventional values of
electronic parameters in the $p$-$d$ model for the CuO$_2$ plane, the CI of two n.n.
holes is estimated by the value $V_{1} = 0.1 - 0.2$~eV.  The  CI for n.n.n. holes, $ 4\,
V_2\, \cos q_x \cos q_y$, is much smaller, $\,V_2/V_1 \sim 0.04\,$ and can be safely
neglected.
\par
As the second model we consider the 2D screened CI   suggested in
Ref.~\cite{Alexandrov11}:
\begin{equation}
V_c({\bf q}) = \frac{2\pi e^2 }{a\, \epsilon_0}
 \frac{1}{a|{\bf q}| +  a\kappa} \equiv u_c \,\frac{1}{a|{\bf q}|
  + a \kappa },
 \label{50}
\end{equation}
where $a$ is the lattice parameter (below we put $a = 1$) and the dielectric constant
$\epsilon_0$ takes into account lattice polarization induced by ligand fields.  We assume
that the screening parameter $\kappa $  depends on the doping and can be described by the
interpolation formula: $ a \kappa = 4 \delta $, so that $\,a \kappa = 0.2$ in  the
underdoped case ($ \delta = 0.05 $) and $a \kappa = 1$ for the overdoped case ($ \delta =
0.25 $).  The energy $u_c $ we estimate from calculation of the CI (\ref{50}) at $\kappa
= 0$ for two n.n. holes at the distance $a_x$ assuming it to be equal to $V_1$ in the
model (\ref{50aa}): $\, V_{c1}(\kappa = 0) =  (u_c / N)\sum_{\bf q}
 (\cos q_x / q ) = V_1\,$. From this equation we get an estimation
$u_c =  V_{1}/0.175 \approx 1$~eV or $\,u_c = 2.5\, t$ for $t = 0.4 $~eV. Here for
convenience, we take $V_ 1 = 0.175$~eV.
\par
In the present study we do not perform  self-consistent computation of spin and charge
excitation spectra  but adopt certain models for the spin (\ref{32a}), charge
(\ref{32b}), and phonon (\ref{32c}) susceptibility  in Eq.~(\ref{31}) or Eq.~(\ref{47}).
Since we consider the electronic spectrum only in the normal state and calculate
superconducting transition temperature $T_c$ from the linearized gap equation (\ref{42})
or (\ref{46}) the feedback effects caused by opening a superconducting gap are not
essential which justifies usage of model functions for the susceptibility.
\par
Due to a large energy scale of charge fluctuations, of the order of several $\,t$,  in
comparison with the spin excitation energy of the order of $J$, the charge fluctuation
contributions in Eq.~(\ref{31}) can be considered in the static limit
\begin{eqnarray}
\chi_{cf}({\bf k}) &=& \chi_{cf,1}({\bf k}) + \chi_{cf,
2}({\bf k}), \label{59} \\
\chi_{cf,1 (2)}({\bf k}) &=& - \frac{1}{N}\sum_{\bf q} \frac{N_{h 1 (2)}({\bf q +k}) -
N_{h 1 (2)}({\bf q})}{\varepsilon_{1 (2)}({\bf q +k}) - \varepsilon_{1 (2)}({\bf q})}.
 \nonumber
\end{eqnarray}
where the  hole occupation numbers $ N_{h 1 (2)}({\bf q})$ are defined in Eq.~(\ref{37}).
It is assumed that the system is far away from a charge instability or a stripe formation
when the energy dependence of the charge susceptibility may be essential (see, e.g.,
Refs.~\cite{Becca96,Castellani98,Gabovich11}).
\par
For the dynamical spin susceptibility $\, \chi_{sf}({\bf q},\omega) = - \langle \langle
{\bf S}_{q}\mid {\bf S}_{-q} \rangle \rangle _{\omega}$ we take a model suggested in
numerical studies~\cite{Jaklic95}
\begin{eqnarray}
  && {\rm Im}\, \chi_{sf}({\bf q},\omega+i0^+) =
 \chi_{sf}({\bf q}) \; \chi_{sf}''(\omega)
 \nonumber \\
& = &\frac {\chi_{ Q}}{1+ \xi^2 [1+ \gamma({\bf q})]} \;  \tanh \frac{\omega}{2T}
\frac{1}{1+(\omega/\omega_{s})^2}\, .
 \label{51}
\end{eqnarray}
The ${\bf q}$-dependence in $\chi_{sf}({\bf q})$ is determined by the  AF correlation
length $\xi$ (in units of $a$).  The frequency dependence is determined by  a broad
spin-fluctuation spectrum  $\chi_{sf}''(\omega)\,$  with  a cut-off energy of the order
of the exchange energy $\omega_s \sim J$. This type of the spin-excitation spectrum was
found also in the microscopic theory for the $t$-$J$ model in Ref.~\cite{Vladimirov09}.
The strength of the spin-fluctuation interaction given by the static susceptibility
$\chi_{ Q} = \chi_{sf}({\bf Q}) $ at the AF wave vector ${\bf Q = (\pi,\pi)}$,
\begin{equation}
\chi_{ Q} = \frac{3 (1- \delta)}{2\omega_{s} }
 \left\{ \frac{1}{N} \sum_{\bf q}
 \frac{1}{ 1+\xi^2[1+\gamma({\bf q})]} \right\}^{-1} ,
 \label{52}
 \end{equation}
is fixed by the normalization condition:
\begin{eqnarray}
&& \frac{1}{N} \sum_{\bf q} \int\limits_{-\infty}^{+\infty}
  \frac{d \omega}{\pi} \, N(\omega)\, {\rm Im}
  \chi_{sf}({\bf q},\omega)=\langle {\bf S}_{i}^2\rangle
  = \frac{3}{4}(1- \delta). \quad
  \nonumber
\end{eqnarray}
Spin correlation functions $C_1,\, C_2$ (\ref{18}) in the single-particle excitation
spectrum (\ref{17}) can be calculated using the same model (\ref{51}):
\begin{equation}
 C_1 = \frac{1}{N} \sum_{\bf q}\, C_{\bf q}\, \gamma({\bf q}), \quad
 C_2 = \frac{1}{N} \sum_{\bf q} \, C_{\bf q}\, \gamma'({\bf q}).
 \label{52a}
\end{equation}
Here the  spin  correlation function $\, C_{\bf q}=\langle {\bf S}_{\bf q}{\bf S}_{-\bf
q} \rangle = C(\xi)/\{1+\xi^2[1+ \gamma({\bf q})]\} \, $ where $C(\xi) = \chi_{ Q}\,
(\omega_{s}/2)$. The results of computation of the correlation functions at several
values of the AF correlation length $\xi$ related to the hole concentrations $\delta$ are
given in Table~\ref{Table1} where the static susceptibility $\chi_{Q}$, the projected
spin susceptibility $\widehat{\chi}\sb{sf}$ (see Eq.~(\ref{57e})) are also given.

\begin{table}
\caption{ Spin correlation functions $C_1,\, C_2$, spin susceptibility $\chi_{ Q}$, and
projected spin susceptibility $\widehat{\chi}\sb{sf}$  for several values of  AF
correlation length $\xi$ related to hole concentration $\delta$. }
 \label{Table1}
\begin{tabular}{crrrrc}
\hline \hline
 $\quad  \xi/a \; $  &   $\; \delta \quad $ &   $ \; C_1 \quad $ &
  $ \; C_2 \quad  $ &  $\; \chi_{ Q}\,\cdot t\quad $ &
   $ \; - \widehat{\chi}\sb{sf} \,\cdot t \; $ \\
\hline
 3.4 & $\;$ 0.05 $\;$ &  - 0.26 $\;$ &   0.16 $\;$ &
29.5  $\;$  & 1.32 $\;$  \\
2.4  & $\;$ 0.10 $\;$ & - 0.20  $\;$  & 0.11 $\;$  &
12.6  $\;$ & 1.05 $\;$  \\
1.5  & $\;$ 0.25  $\;$ & - 0.12 $\;$  &  0.05 $\;$  &
6.8 $ \;$ & 0.61 $\;$  \\
\hline \hline
\end{tabular}
\end{table}
\par

To estimate the contribution from phonons in Eq.~(\ref{31}) we consider  a model
susceptibility for optic phonons and the EPI matrix element in the form similar to
Ref.~\cite{Lichtenstein95}:
\begin{equation}
V_{ep}({\bf q},\omega)= |g({\bf q})|^2\chi_{ph}({\bf q},\omega) = g_{ep}\,
\frac{\omega_0^2}{\omega^2_{0} - \omega^2}\,S(q),
   \label{53}
\end{equation}
where $ g_{ep}\,$ is the ``bare''  matrix element for the short-range  EPI, while the
momentum dependence of the EPI is determined by the  vertex correction $ S(q)\,$.  It
takes into account a strong suppression of charge fluctuations at small distances (large
scattering momenta $q$) induced by electron correlations as proposed in
Ref.~\cite{Zeyher96}. For the vertex function we take the model
\begin{equation}
 S(q)= \frac{1}{\kappa_1^2 + q^2}\equiv
 \frac{\xi_{ch}^2}{1 + \xi_{ch}^2 \,q^2},
    \label{53a}
\end{equation}
where the charge correlation length $\xi_{ch} = 1/\kappa_1\,$ determines the radius of a
``correlation hole''. Taking into account that $\xi_{ch} \sim a/\delta$~\cite{Zeyher96},
we can use the relation $\xi_{ch} = 1/ (2 \delta )$ in numerical computations. This gives
$\xi_{ch} \simeq 10$ for the underdoped case ($ \delta = 0.05 $) and $\xi_{ch}\simeq 2$
for the overdoped case ($ \delta = 0.25 $). We assume a strong   EPI $g_{ep} = 5\, t =
2.0$~eV and take  $\omega_{0} = 0.1\,t = 0.04$~eV.
\par
In computations we use the following parameters for the model (\ref{2}): $\, U =
\Delta_{pd} = 8 \, t, \quad t' = - 0.2 \, t, \quad t'' = 0.10 \, t$. As an energy unit we
use $t = 0.4$~eV. The exchange interaction  is described by the function $\,J({\bf q})= 2
J\, (\cos q_x + \cos q_y ) $ with $J = 0.4t$. For the CI energy for the n.n. holes we
take    $V_1 = 0.44 t$ and $u_c = 2.5 t$.  The electronic spectrum in the normal state is
calculated at  $T = 0.02t \sim 100$~K.  In computations the grid of $64 \times 64$ $(k_x,
k_y)$ points   and up to 1200 imaginary frequencies $\omega_n$ were used.

\subsection{Electronic spectrum in the normal state}
\label{sec:8}

\begin{figure}
\resizebox{0.35\textwidth}{!}{%
\includegraphics{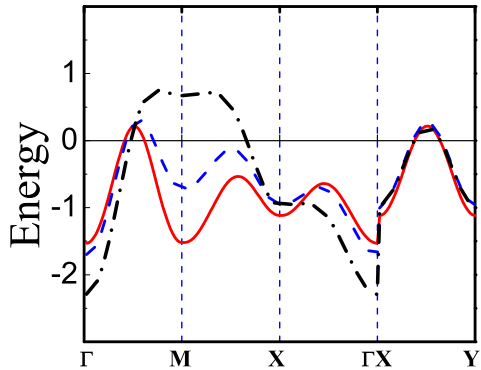}}
 \caption{(Color online) Electron dispersion in the MFA
 ${\varepsilon}_{2} ({\bf k})$  along the symmetry directions $\Gamma(0, 0)\rightarrow
M(\pi,\pi) \rightarrow X (\pi, 0) \rightarrow \Gamma(0, 0)$ and $X (\pi, 0) \rightarrow
Y(0, \pi)$ for $\delta = 0.05\,$ (red solid line), $0.10$ (blue dashed  line), and $0.25$
(black dash-dotted line). Fermi energy for hole doping is at $\omega = 0$.}
 \label{fig1p}
\end{figure}
At first we consider  results in the MFA for the electronic spectrum (\ref{17}). The
doping dependence of the electron dispersion for the two-hole subband  ${\varepsilon}_{2}
({\bf k})$  along the symmetry directions in the 2D Brillouin zone (BZ) is shown in
Fig.~\ref{fig1p}. For small doping, $\, \delta = 0.05$, the energy at the $M(\pi,\pi)$
and $\Gamma(0, 0)$ points are nearly equal as in the AF long-range order state. Only
small hole-like FS pockets close to the $(\pm \pi/2,\pm \pi/2)$ points emerge at this
doping. With increasing doping, the AF correlation length decreases that results in
increasing  of the electron energy at the $M(\pi,\pi)$ point and at some critical doping
$\delta \sim 0.12$ a large FS appears. At the same time,  the renormalized two-hole
subband width increases with doping from $\widetilde{W} \approx 2\,t$ at $\, \delta =
0.05\,$ to $\widetilde{W} \approx 3\,t$ at $\, \delta = 0.25$, which, however, remains
less than the ``bare'' Hubbard bandwidth $W = 4 t\,(1+\delta)$ where short-range  AF
correlations are disregarded. Note that in the dynamical mean field theory (DMFT) this
narrowing of the subbands due to the short-range AF correlations is
missed~\cite{Georges96,Kotliar06}, while they are taken into account partly in the
cluster DMFT~\cite{Haule07}. However, as shown in the DMFT  the self-energy contribution
strongly renormalizes the electronic spectrum found in the MFA.
\par
\begin{figure}
\centering
\resizebox{0.35\textwidth}{!}{%
\includegraphics{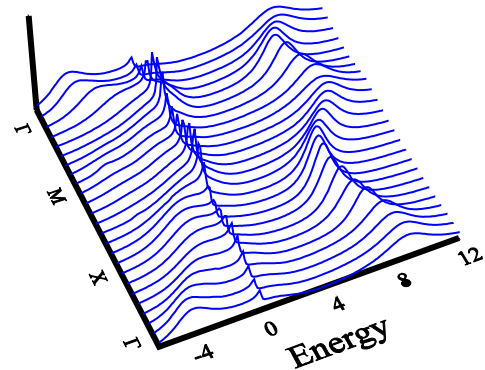}}
\caption{Spectral function in the SCA along the symmetry directions $\Gamma(0,
0)\rightarrow M(\pi,\pi) \rightarrow X (\pi, 0) \rightarrow \Gamma(0, 0)$ for hole
concentration $\delta = 0.10$.}
 \label{fig4p}
\end{figure}
\begin{figure}
\centering
\resizebox{0.35\textwidth}{!}{%
\includegraphics{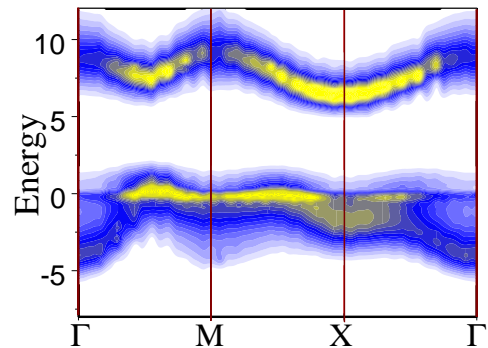}}
\caption{(Color online) Electron dispersion curves in the SCA along the symmetry
directions $\Gamma(0, 0)\rightarrow M(\pi,\pi) \rightarrow X (\pi, 0) \rightarrow
\Gamma(0, 0)$ for hole concentration $\delta = 0.10$. }
 \label{fig5p}
\end{figure}
\begin{figure}
\centering
\resizebox{0.35\textwidth}{!}{%
\includegraphics{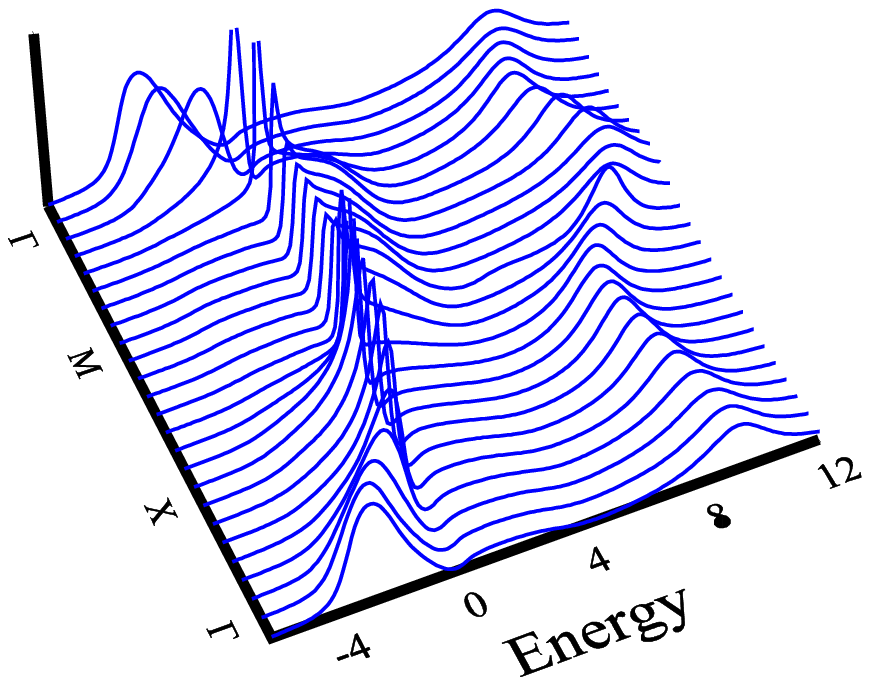}}
\caption{The same as in Fig.~\ref{fig4p} for  hole concentration $\delta = 0.25$.}
 \label{fig6p}
\end{figure}
\begin{figure}
\centering
\resizebox{0.35\textwidth}{!}{%
\includegraphics{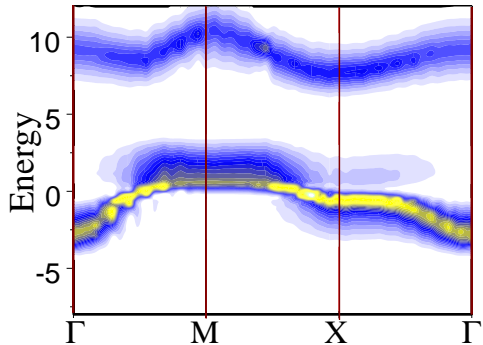}}
\caption{The same as in Fig.~\ref{fig5p} for  hole concentration $\delta = 0.25$.}
 \label{fig7p}
\end{figure}

To consider the self-energy effects in  the electronic spectrum a strong coupling
approximation (SCA) should be considered by  a self-consistent solution of the system of
equations for the normal GF (\ref{34}) and the self-energy (\ref{35}). In
Ref.~\cite{Plakida07} a detailed investigation of the normal state electronic spectrum
for the conventional Hubbard model in SCA  was performed. Therefore, here we present only
the  results of the electronic spectrum computation for the  model (\ref{2}) which are
important for further studies of superconductivity in the model. The  spectral functions
(\ref{40}) along the symmetry directions are presented in Figs.~\ref{fig4p}
and~\ref{fig6p} for $\delta = 0.10$ and $\delta = 0.25$, respectively. The dispersion
curves given by the maximum of the spectral function  (\ref{40}) at the same doping are
displayed in Figs.~\ref{fig5p} and~\ref{fig7p}.

In comparison with  the MFA in Fig.~\ref{fig1p}, a rather flat energy dispersion is found
with QP peaks at the FS. In general, strong increase of the dispersion and  intensity of
the QP peaks is observed in the overdoped region in  comparison with the underdoped
region. This is in agreement with our detailed studies of temperature and doping
dependence of the self-energy (\ref{35}) and spectral function (\ref{40})
in~\cite{Plakida07} which have proved  strong influence of AF spin-correlations on the
spectra.
\par
\begin{figure}[!]
\centering
\resizebox{0.35\textwidth}{!}{%
\includegraphics{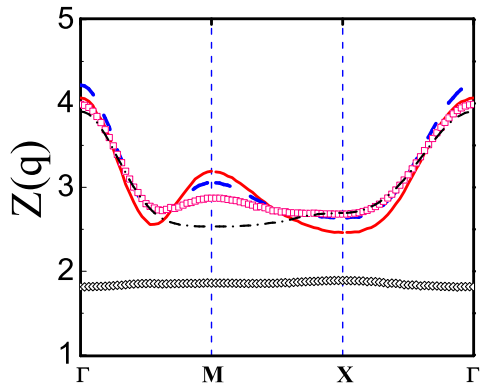}}
\caption{(Color online) Doping dependence of the renormalization parameter $Z({\bf q}) $
along the symmetry directions $\Gamma(0, 0)\rightarrow M(\pi,\pi) \rightarrow X (\pi, 0)
\rightarrow \Gamma(0, 0)$ at $T \approx 140$~K for $\delta = 0.05$  (red solid line),
$\delta = 0.10$ (blue dashed line), $\delta = 0.15$ (pink squares), $\delta = 0.25$
(black dash-dotted line), and $\delta =  0.35$ (black diamonds).}
 \label{fig8p}
\end{figure}
\begin{figure}[!]
\centering
\resizebox{0.35\textwidth}{!}{%
\includegraphics{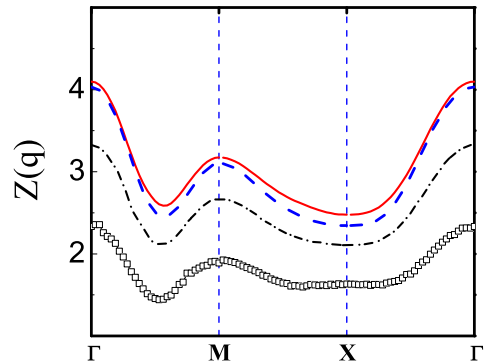}}
\caption{(Color online) Temperature  dependence of the renormalization parameter $Z({\bf
q}) $ for $\delta = 0.05$  at $T \approx 140$~K (red solid line),  $T \approx 580$~K
(black dash-dotted line), and $T \approx 1100$~K  (black  squares). Blue dashed line
shows $Z({\bf q}) $ for $\delta = 0.05$   caused only by the spin-fluctuation
contribution.}
 \label{fig9p}
\end{figure}
\begin{figure}[!]
\centering
\resizebox{0.35\textwidth}{!}{%
\includegraphics{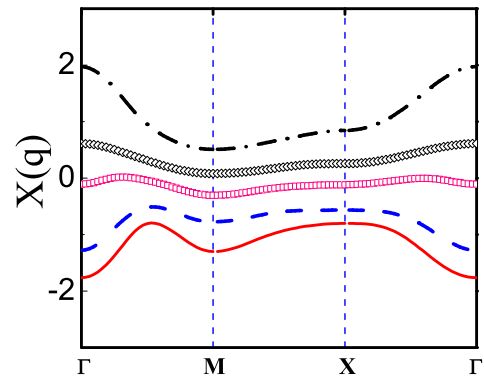}}
\caption{(Color online) Doping dependence of the parameter $X({\bf q}) $. Notation are
the same as in Fig.~\ref{fig8p}.}
 \label{fig10p}
\end{figure}

To estimate the coupling constant $\lambda({\bf q})$ in the two-hole subband, we
calculated the renormalization parameter $Z({\bf q})$ (\ref{45b}) at the Fermi energy,
 \begin{eqnarray}
 Z({\bf q})& = &Z({\bf q}, \omega =0)= 1 + \lambda({\bf q})
 \nonumber \\
 & =&    1 - [\, d \,{\rm Re}\,
  \Sigma({\bf q}, \omega)/{d \omega}]|_{\omega = 0}.
 \label{45ba}
 \end{eqnarray}
The doping dependence of  $Z({\bf q}) $ is shown in Fig.~\ref{fig8p}. It weakly depends
on $\delta$ in the underdoped case for $\delta \lesssim 0.15$ but sharply decreases in
the overdoped case for $\delta \gtrsim 0.25$. The temperature dependence of  $Z({\bf q})
$ presented  in Fig.~\ref{fig9p} is weak  at temperatures lower than  the characteristic
energy of spin fluctuations $\omega_s \sim J$. The EPI gives a small contribution to the
coupling constant as follows from the comparison of $Z({\bf q}) $ induced by both
spin-fluctuations and EPI contributions (red solid line in Fig.~\ref{fig9p})  with the
contribution caused only  by spin-fluctuations  (blue dashed line).
The renormalization parameter $X({\bf q})$ (\ref{45c}) at the Fermi energy
\begin{eqnarray}
 X({\bf q}) =  X({\bf q},\omega = 0)
= {\rm Re}{\Sigma}({\bf q}, 0) ,
 \label{45ca}
 \end{eqnarray}
which determines the shift of the dispersion curve is plotted along the symmetry
directions in Fig.~\ref{fig10p}. $X({\bf q}) $ decreases with doping in the underdoped
region as $Z({\bf q})$, but in the overdoped region reveals an irregular behavior and
becomes small  at large doping as $Z({\bf q})$. These results demonstrate that at large
doping both the electron interaction with  spin-fluctuations and the EPI become weak.

\subsection{Superconducting gap and $ \bf T_c$}
\label{sec:9}

For a comparison of various contributions  to the superconducting gap equation
(\ref{42}), as the first step, we consider a weak-coupling approximation (WCA). In the
WCA, the interaction (\ref{31}) is approximated by its value close to the Fermi energy,
$\, |\omega, z | \sim 0$.  Then integration over $\Omega$ of the dynamical susceptibility
in (\ref{31}) yields
\begin{eqnarray}
\int\limits\sb{-\infty}\sp{+\infty}
   \frac{d\Omega}{\pi}\,\frac{{\rm Im}\,\chi({\bf q},\Omega)}
   {\omega - z - \Omega} \simeq \int\limits\sb{-\infty}\sp{+\infty}
 \frac{d\Omega}{\pi}\,\frac{{\rm Im}\,\chi({\bf q},\Omega)}
   { - \Omega}
     = - \chi({\bf q}), \quad
    \label{54}
\end{eqnarray}
where   $\chi({\bf q}) ={\rm Re} \, \chi({\bf  q},\Omega = 0) $ is the static
susceptibility. In the WCA the self-energy contribution in the normal-state GF (\ref{34})
is neglected that results in the BCS-type equation for the gap function   at the Fermi
energy  $\varphi({\bf k}) = \sigma\,\varphi_{2, \sigma}({\bf k},\omega =0)$:
\begin{eqnarray}
&& \varphi({\bf k})=
 \frac{1}{N} \sum_{\bf q}[1-b({\bf q})]^2
 \frac{\varphi({\bf q})}{2 E({\bf q})} \tanh \frac{E({\bf q})}{2T}
 \big\{ J({\bf k-q})
\nonumber \\
&& - V({\bf k-q})+ \big[(1/4) |t({\bf q})|\sp{2} + |V({\bf k -q})|^2\big]
\chi\sb{cf}({\bf k -q})
 \nonumber \\
&& + |g({\bf k -q})|^2 \, \chi_{ph}({\bf k-q})\,\theta(\omega_{0} -| \varepsilon_{2}({\bf
q})|)
\nonumber \\
&& -  |t({\bf q})|\sp{2}\; \chi_{sf}({\bf k -q})\theta(\omega_{s} -| \varepsilon_{2}({\bf
q})|) \big\}\,,
     \label{55}
\end{eqnarray}
where $ E({\bf q}) = [\varepsilon^2_{2}({\bf q}) + |\varphi({\bf q})|^2]^{1/2}$.  Whereas
for the exchange interaction and CI there are no retardation effects and the pairing
occurs for  all electrons in the two-hole subband, the EPI and spin-fluctuation
contributions  are restricted  to the range of  energies $\,\pm \omega_0$ and $\,\pm
\omega_s$, respectively, near the FS, as determined by the $\theta$-functions.
\par
To estimate various contributions in the gap equation (\ref{55}) we consider a model
$d$-wave gap function, $\varphi({\bf k}) = (\Delta/2) \,\eta({\bf k})$ where $ \eta({\bf
k}) = (\cos k_x - \cos k_y)$. Then the  gap equation  can be written in the form:
\begin{eqnarray}
&& 1 = \frac{1}{N } \sum_{\bf q}[1-b({\bf q})]^2
 \frac{\eta({\bf q})^2} {2  E_{\bf q}}
  \tanh\frac{ E_{\bf q}}{2T}\big\{ J - \widehat{V_c}
 \nonumber \\
& & +  \widehat{V}\sb{cf} + (1/4)\,|t({\bf q})|\sp{2} \widehat{\chi}\sb{cf} +
\widehat{V}\sb{ep}\,
   \theta(\omega_{0} -| \varepsilon_{2}({\bf q})|)
 \nonumber \\
& &
   - |t({\bf q})|^{2}\, \widehat{\chi}\sb{sf}
   \theta(\omega_{s} -| \varepsilon_{2}({\bf q})|) \big\}.
\label{56}
\end{eqnarray}
In this equation only $l=2$ components of the static susceptibility  and CI give
contributions
\begin{eqnarray}
 \widehat{V_c} & = & \frac{1}{N}\sum_{\bf k} V({\bf k})\cos k_x,
\label{57a}\\
 \widehat{V}\sb{cf} & = & \frac{1}{N}\sum_{\bf k}
   |V({\bf k })|^2\chi\sb{cf}({\bf k})\,\cos k_x,
\label{57b} \\
\widehat{\chi}\sb{cf} & = &\frac{1}{N}\sum_{\bf k}
 \chi\sb{cf}({\bf k}) \cos k_x ,
\label{57c} \\
 \widehat{V}_{ep} & = &  \frac{g_{ep}}{N}\sum_{\bf k}
 S({\bf k})\cos k_x,
 \label{57d} \\
  \widehat{\chi}\sb{sf} & = &  \frac{1}{N}\sum_{\bf k}
 \chi\sb{sf}({\bf k})\cos k_x \, .
\label{57e}
\end{eqnarray}
Computation yields the following parameters for the n.n. intersite CI (\ref{50aa}):
$\widehat{V_c} = V_1 = 0.44 t \approx 0.18$~eV. For the screened CI (\ref{50}) we have:
\begin{eqnarray}
\widehat{V_c}(\kappa ) &= &  \frac{u_c}{N}\sum_{\bf q}
 \frac{\cos q_x}{q +\kappa},
  \label{58}
\end{eqnarray}
where $\widehat{V_c}(\kappa) = 0.12 \,t \; (0.28\, t) \approx 0.05\; (0.11)$~eV for
$\kappa = 1 \, ( 0.2)\,$, respectively (see Table~\ref{Table2}). Note, that the projected
CI (\ref{58}) is much smaller than the CI energy $\,V_{c0}(\kappa) = (u_c / N)\sum_{\bf
q} ([1/ ( q + \kappa)] $. In particular, $\widehat{V_c}(\kappa)/V_{c0}(\kappa) = 0.15
\; (0.24 )$ for $\kappa = 1, \,(0.2)$, respectively. In the conventional BCS theory the
CI is suppressed by retardation effects described  by large Bogoliubov-Morel logarithm,
$\ln(\mu/\omega_{ph})$. In the Hubbard model there are no retardation effects for the AF
exchange interaction but a reduction  of the CI contribution is due to the
$d$-wave pairing.
\par
To estimate contributions from  the charge fluctuations  we use the static charge
susceptibility (\ref{59}). Applying this approximation to the screened CI (66) we get the
following expression for charge contribution (79):
\begin{eqnarray}
 \widehat{V}_{cf}(\kappa)  & = &\frac{u_c^2 }{N} \sum_{\bf k}
 \,\frac{1}{(k
  + \kappa)^2 } \, \chi\sb{cf}({\bf k}) \, \cos k_x,
  \label{60}
\end{eqnarray}
where $\, \widehat{V}_{cf}(\kappa) = 0.05 \; ( 0.25 )\, t \approx 0.02\;( 0.1 )$~eV for $
\kappa = 1\, ( 0.2)$, respectively. This contribution is smaller  than the CI repulsion
$\widehat{V_c} $ (\ref{58}) and in our approximation the $d$-wave pairing induced by the
screened CI in the second order $ \widehat{V}_{cf}$ is destroyed by CI repulsion in the
first order  $\widehat{V_c} $ as was pointed out in Ref.~\cite{Alexandrov11}. The charge
fluctuation contribution from the   n.n. intersite CI (\ref{50aa}) is even smaller, $\;
\widehat{V}_{cf}^{nn} \approx 4 \cdot 10^{-3}\, t\; $. The contribution from the charge
fluctuations (\ref{57c}) calculated for the static susceptibility (\ref{59}) is also
small: $\, \widehat{\chi}\sb{cf}(\delta) = (1/t)\,0.15 \cdot 10^{-2}\;( 1.3  \cdot
10^{-2}) $ for the hole concentrations $\delta = 0.05 \; ( 0.10 )$, respectively. For the
averaged over the BZ  vertex $ \overline{|t({\bf q})|^2} = (1/N)\sum_{\bf q}|t({\bf
q})|^2 \simeq 4\,t^2 $ this contribution is equal to $\overline{|t({\bf
q})|^2}\,\widehat{\chi}\sb{cf} \lesssim 0.02$~eV and can be neglected.
\par
The EPI contribution (\ref{57d})  is given by
\begin{equation}
\widehat{V}_{ep} =  \frac{g_{ep}}{N}\sum_{\bf k}\frac{\xi_{ch}^2}{1 + \xi_{ch}^2 \,k^2}
 \cos k_x
 \equiv g_{ep} \, S_d(\xi_{ch}),
 \label{61}
\end{equation}
where $S_d(\xi_{ch})= 0.154\, (0.393)$ for $\xi_{ch} = 2 \, (10)$, respectively. Thus,
even for a strong EPI coupling $g_{ep} = 5 t = 2$~eV we obtain a moderate contribution
from the EPI for the $d$-wave pairing: $\widehat{V}_{ep}(\xi_{ch})= 0.76\, t\, (1.96 \,t)
\approx 0.3\, (0.8)$~eV for $\,\xi_{ch} = 2\, (10)$, respectively. The EPI contribution
to the $s$-wave pairing is given by the $l = 0$ component $ S_0 =(1/N)\sum_{\bf q}\,
S(q)= 0.31\; (0.57)\,$ for $\xi_{ch} = 2\; (10)$, respectively. The ratio of the $d$-wave
$S_d$ and the $s$-wave $S_s = S_0$ components of the EPI matrix elements is equal to
$\,(S_d/ S_0) =  0.43\; ( 0.60)$ for $\xi_{ch} =2 \; (10)$, respectively. This shows that
at small hole concentrations $\delta$ (large charge correlation lengths $\xi_{ch} = 1/
2\delta $)  the EPI for the both components are comparable, while for the overdoped case
the $d$-wave component $S_d$ becomes considerably smaller than the $s$-wave component in
agreement with the results of Ref.~\cite{Zeyher96}.
 \par
The spin-fluctuation contribution $\widehat{\chi}\sb{sf}$ calculated for the model
$\chi_{sf}({\bf q})$ in Eq.~(\ref{51}) for several values of the AF correlation length
$\xi$   is given in Table~\ref{Table1}. Using the averaged over  BZ  vertex $\,
\overline{|t({\bf q})|^2} \simeq 4\,t^2\,$ we can estimate an effective spin-fluctuation
coupling constant as $\, g_{sf} \simeq - 4\,t^2 \,\widehat{\chi}\sb{sf} = 5.3\, (2.4)t
\approx 2\, (1)$~eV for $\delta = 0.05\, (0.25)$,  respectively.  Thus,  the
spin-fluctuation contribution to the pairing in Eq.~(\ref{56}) appears to be  the
largest. Note, that  $\, g_{sf}\,$ is close to the spin-fluctuation coupling constant
$\widetilde{U} \approx 1.6$~eV found in Ref.~\cite{Dahm09} from ARPES data.
\par\begin{table}
\caption{ CI parameters  $\widehat{V}_{c},\, {V}_{c0}, \; \widehat{V}_{cf}$, and EPI
parameter $\widehat{V_{ep}}$ for several values of the CI screening constant $\kappa = 4
\delta$ and the charge correlation length $\xi_{cf}  = 1/ 2\delta $ for EPI related to
hole concentration $\delta$. }
 \label{Table2}
\begin{tabular}{crrrrrc}
\hline \hline $\; \delta \quad $  &   $\; \kappa \quad $  &   $ \; \xi_{cf} \; $ &
  $ \;\widehat{V}_{c}\,/ t \quad  $ &$ \;{V}_{c0}\,/ t \quad  $&
 $\; \widehat{V}_{cf}\,/t \quad $ & $ \;  \widehat{V_{ep}} \,/ t \; $ \\
\hline $\;$ 0.05 $\;$ & 0.2 $\;$ &  10 $\;$ & 0.28 $\;$ & 1.18
$\;$& 0.25 $\;$  & 1.96 $\;$  \\
$\;$ 0.10 $\;$   & 0.4 $\;$ & 5  $\;$  & 0.22 $\;$  & 1.05 $\;$  &
0.26  $\;$ & 1.4 $\;$  \\
$\;$ 0.25  $\;$   & 1 $\;$ & 2 $\;$  &  0.12 $\;$  &  0.80 $\;$ &
 $0.05 \;$ & 0.76 $\;$  \\
\hline \hline
\end{tabular}
\end{table}
In Table~\ref{Table2}  we present  the coupling parameters in the equation for $T_c$
(\ref{56}). In the  MFA the pairing can be induced by the AF exchange interaction  $J=
0.4 t = 0.16$~eV which is comparable with the repulsion caused by the screened CI:
$\widehat{V_c} = (0.05 - 0.11)$~eV or even smaller than the n.n. hole CI $V_1 =
0.175$~eV. Therefore, the superconducting pairing in the  MFA for the $t$-$J$ model (in
particular, the  RVB state ~\cite{Anderson87}) is strongly suppressed (or even destroyed)
by the intersite Coulomb repulsion.
\par
\begin{figure}
\centering
\resizebox{0.35\textwidth}{!}{%
\includegraphics{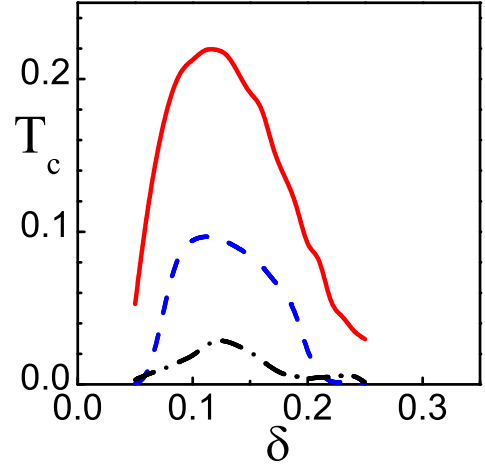}}
\caption{(Color online) $T_c(\delta)$ in the WCA induced by all interactions (red solid
line) and only by the spin-fluctuation contribution $\widehat{\chi}\sb{sf}$(blue dashed
line) or only by the EPI $\, \widehat{V}\sb{ep}$ (black dash-dotted line). }
 \label{fig14p}
\end{figure}

To calculate doping dependence of $T_c$ we solve Eq.~(\ref{56}) by taking into account
the exchange interaction $J$, the Coulomb repulsion $\widehat{V_c}$, and the
contributions from the self-energy in the WCA:  $\, \widehat{V}\sb{ep}$,
$\widehat{\chi}\sb{sf}$, and  $\, \widehat{V}\sb{cf} $, neglecting the small contribution
$\widehat{\chi}\sb{cf}$. Results of the calculation is shown in Fig.~\ref{fig14p}.  The
highest $T_c \approx 0.2\, t $ is found when all the contributions are taken into
account. The spin-fluctuation pairing results in superconducting  $T^{sf}_c \approx
0.04\, t $  much larger than $T^{ep}_c \approx 0.01 t$ mediated by the EPI. For the ${\bf
k}$-independent EPI ($S({\bf k}) =1$) as in the Holstein model, the $d$-wave pairing is
absent. The doping dependence of $T_c$ is qualitatively agree with experiments in
cuprates but its value is an order of magnitude higher.

\begin{figure}
\centering
\resizebox{0.35\textwidth}{!}{%
\includegraphics{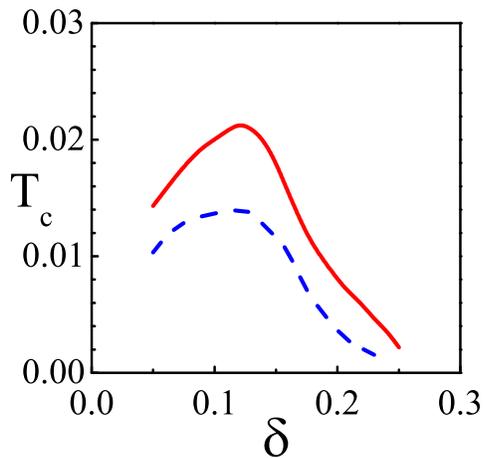}}
\caption{(Color online)  $T_c(\delta)$ induced by all interactions in Eq.~(\ref{64}) (red
solid line) and only by spin-fluctuation contribution $V_{sf}$ (blue dashed line).}
 \label{fig15p}
\end{figure}

The high values for $T_c$ found in  the WCA are explained by neglecting the reduction of
the quasiparticle weight caused by the self-energy effects in the gap equation
(\ref{46}).  It is convenient to  write the gap equation in the form:
\begin{eqnarray}
&&\varphi({\bf k},\omega_n) =
  \frac{T_c}{N}\sum_{\bf q} \,  \sum_{m}\,
\big\{ J({\bf k -q}) - V_c({\bf k -q})
\nonumber \\
& &  - V_{sf}({\bf q, k-q},\omega_{n}- \omega_{m})
 + V_{ep}({\bf k-q},\omega_{n}- \omega_{m})\big\}
\nonumber\\
&&\times  \frac{[1 - b({\bf q})]^2\,
  \varphi({\bf q}, \omega_{m})}{[\omega_m Z({\bf q},\omega_m)]^2
   + [{\varepsilon}_{2}({\bf q})+ X({\bf q},\omega_m)]^2}\, .
 \label{64}
\end{eqnarray}
For  $ V_c({\bf k -q})$ we take the screened CI (\ref{50}). Since the charge-fluctuations
gives a much weaker  contribution  than the spin-fluctuation and electron-phonon
interactions (see Table~\ref{Table1} and Table~\ref{Table2}), we neglect the term
$[|V({\bf k -q})|^2 + |t({\bf q})|^{2}/4 ] \, \chi_{cf}({\bf k-q}, \nu_n)$ in the
interaction function(\ref{47}). Contributions induced by spin-fluctuations and the EPI
are described by the functions
\begin{eqnarray}
&& V_{sf}({\bf q, k-q}, \omega_{\nu})=  |t({\bf q})|^{2}\,
 \chi_{sf}({\bf k-q})
  F_{sf}(\omega_{\nu}),\quad
 \label{65a} \\
&& V_{ep}({\bf  k-q}, \omega_{\nu})= g_{ep}\, \frac{\omega_0^2}{\omega^2_{0} +
\omega_\nu^2}\,S({\bf k - q}),
  \label{65b}
\end{eqnarray}
where  the spectral function for spin fluctuations reads:
\begin{equation}
F_{sf}(\omega_\nu)=\frac{1}{\pi} \int_{0}^{\infty}\frac {2x dx}{x^2 +
(\omega_\nu/\omega_s)^2}\frac{\tanh ( x\,\omega_{s}/ 2T)}{1+x^2} \, .
 \label{65a1}
\end{equation}
To calculate $T_c$ and to find out the energy- and ${\bf k}$-dependence of the gap $\,
\varphi({\bf k},\omega)$, Eq.~(\ref{64}) was solved by a direct diagonalization in $(k_x,
k_y,\, \omega_n)$-space. Since the largest contribution in Eq.~(\ref{64}) comes from
energies close to the FS, we have used the renormalization parameters at the Fermi energy
$Z({\bf q})$ (\ref{45ba}) and $X({\bf q}) $ (\ref{45ca}) instead of the energy dependent
ones. The results for $T_c(\delta)$ is shown in Fig.~\ref{fig15p}. The highest $T_c \sim
0.021 t \sim 100$~K   is found when all the contributions are taken into account, though
pairing induced only by spin-fluctuations also results  in  high $T^{sf}_c \sim 0.014 t
\sim 65$~K. The $d$-wave pairing induced only by the EPI is rather weak and does not
displayed in Fig.~\ref{fig15p}. The value of $T_c$ is  reduced by an order of magnitude
in comparison with the WCA in Fig.~\ref{fig14p} due to a suppression of the QP weight by
the factor $[1/Z({\bf q})]^2$. The maximum value of $T_c$ is found  at lower value of
doping $\delta_{opt} \approx 0.12$ than in experiments, $\delta^{exp}_{opt} = 0.16$.
\par
\begin{figure}
\centering
\resizebox{0.35\textwidth}{!}{%
\includegraphics{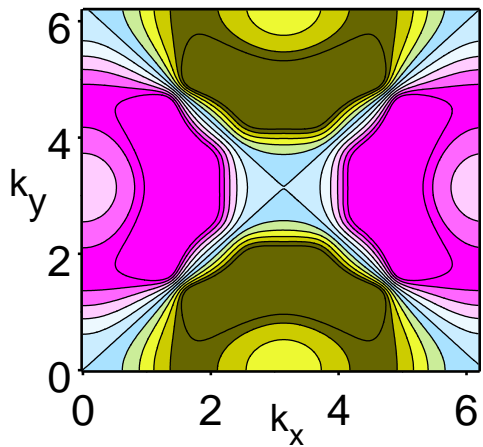}}
\caption{(Color online) 2D plot of the SC gap $\varphi({\bf k},\omega \simeq 0)$.}
 \label{fig16p}
\end{figure}
\begin{figure}
\centering
\resizebox{0.35\textwidth}{!}{%
\includegraphics{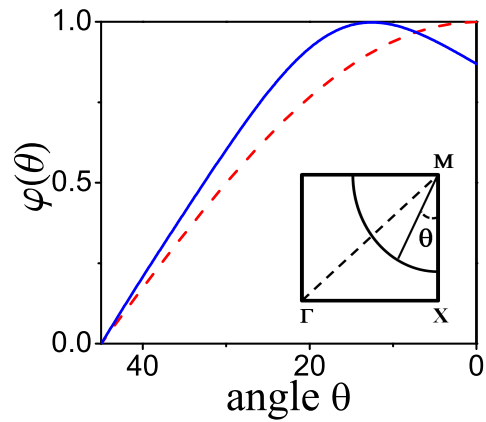}}
\caption{(Color online) Angle dependence of the SC gap $\varphi(\theta) $ on the FS (blue
bold line) in comparison with the model $d$-wave dependence $\varphi_d(\theta) = \cos
2\theta$ (red  dashed lines).}
 \label{fig17p}
\end{figure}
\begin{figure}
\centering
\resizebox{0.35\textwidth}{!}{%
\includegraphics{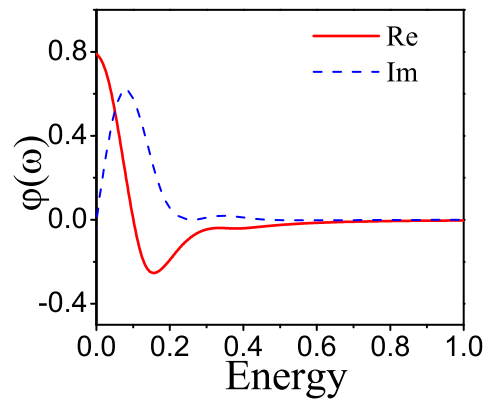}}
\caption{Energy  dependence of the real, Re~$\varphi({\bf k},\omega)$, and imaginary,
Im~$\varphi({\bf k},\omega)$, parts of  the SC gap in arbitrary units.}
 \label{fig18p}
\end{figure}

The ${\bf k}$-dependence of the gap function  $\varphi({\bf k}, \omega \simeq 0)\,$ at
doping $\delta = 0.13$ for $\, (0 \leq k_x,\, k_y \leq 2\pi)$ is plotted in
Fig.~\ref{fig16p}. The gap reveals a distinct $d$-wave symmetry with maximum values in
the vicinity of  the FS.  As shown in Fig.~\ref{fig17p}, its  angle dependence on the FS
is close to the model $d$-wave dependence $\varphi_d(\theta) = \cos 2\theta$.
Energy dependence  (in units of $\,t \,$) of  the gap function $\varphi({\bf k},\omega)$,
the real and imaginary parts, is presented in Fig.~\ref{fig18p} at ${\bf k}\approx
(0,\pi/2)$ and $\delta = 0.13$. Since the gap function was obtained as a solution of the
linear equation at $T = T_c$ the  value of the gap is given in arbitrary units. The
energy  variation of the gap occurs in the region of  $\omega \lesssim 0.4 t $, of the
order of the characteristic  spin-fluctuation energy $\omega_s = J = 0.4 t$.

Generally, the  results obtained in the SCA are in a qualitative agreement with
experiments in the cuprate superconductors. They also demonstrate an important role of
the self-energy effects in the normal and superconducting states in comparison with the
MFA.

\subsection{Comparison with previous theoretical studies}

As briefly discussed in Sec.~\ref{sec:1}, various  methods have been used in theoretical
studies of superconductivity in the Hubbard model. Here we would like to emphasize  our
results in comparison with  previous investigations of the problem.
\par
At first we refer to  results obtained in  the weak or intermediate  correlation limit.
In particular, using  the two-particle self-consistent non-perturbative approach (see,
e.g. Refs.~\cite{Vilk95,Tremblay06}) and the fluctuation-exchange (FLEX) approximation
(see, e.g., Refs.~\cite{Bickers89,Monthoux94a} and reviews~\cite{Moriya00,Manske04}), a
system of equations was derived within the Fermi-liquid model to study self-consistently
single-electron GFs and the spin and charge dynamical susceptibility. Within the FLEX
approximation,  the superconducting $d$-wave pairing was found in a narrow range of
doping, very close to the AF instability. In our theory superconductivity is mediated by
a broad spectrum of AF spin excitations (paramagnons) which results in the doping
dependence of $T_c$ close to experimentally observed  (see Figs.~\ref{fig14p},
\ref{fig15p}).
\par
A general problem of the weak CI in the Hubbard model has been extensively studied within
the RG approach (for reviews see Refs.~\cite{{Shankar94},Metzner98}). The RG studies
revealed a competition between various type of phases driven by electronic instability,
such as the spin-density wave (SDW), charge-density wave (CDW), nematic (Pomeranchuk)
phase, stripes, superconducting pairing, etc. (see  reviews ~\cite{Kivelson03,Vojta09}
and Refs.~\cite{Furukawa98,Honer01,Honer07,Halb00a,Halb00b}). For a weak Hubbard
interaction $U \sim t$ in a certain range of hopping parameters and doping the $d$-wave
superconducting pairing can overcome other instabilities. An important role of the
intersite Coulomb repulsion in the Hubbard model was found in Ref.~\cite{Raghu12}, as
mentioned in Sec.~\ref{sec:1}. In our theory we  disregarded other instabilities and did
not study a general phase diagram since it would demand investigation of a much more
complicated system of equations which is beyond the scope of the present paper.
\par
To consider cuprate superconductors,  the strong correlation limit should be
investigated. In many publications the spectrum of electronic excitations in the normal
state in the Hubbard model was extensively studied.  Here we refer to numerical
simulations for finite clusters (see reviews~\cite{Dagotto94,Bulut02,Scalapino07}), the
DMFT (see reviews~\cite{Georges96,Kotliar06}), the dynamical cluster approximation
(DCA)~\cite{Maier05,Maier06} and the cluster DMFT (see, e.g.,
Refs.~\cite{Haule07,Kancharla08}).  More accurate results have been obtained within the
DCA and cluster DMFT methods where short-range AF correlations are partially taken into
account. As we  have pointed out in Sec.~\ref{sec:4a} and Sec.~\ref{sec:8}, in our method
short-range AF correlations are properly taken into consideration in the MFA resulting in
a large reduction of the effective bandwidth $\,\widetilde{W}$. Consequently, a
two-subband state in the Hubbard model is found even in the intermediate correlation
limit, $U \sim 4t $. The spectral density computed in SCA, Figs.~\ref{fig4p}
and~\ref{fig6p}, are in accord with numerical studies for the Hubbard model (see, e.g.,
Refs.~\cite{Maier05,Tremblay06,Avella07}).
\par
The most controversial problem is whether the superconductivity can emerge from the
repulsion, as discussed in Sec.~\ref{sec:1}. Extensive numerical studies for finite
clusters have revealed a tendency to the $d$-wave pairing in the Hubbard model, though a
delicate balance between superconductivity and other instabilities (AF, SDW, CDW, etc.)
was found (see, e.g., Refs.~\cite{Bulut02,Scalapino07,Maier05,Maier06,Kancharla08}). In
Ref.~\cite{Macridin09}),  using the DCA with the quantum Monte Carlo method, the
superconducting $d$-wave pairing and the isotope effect similar to  observed in cuprates
were found for the Hubbard-Holstein model. However, in several publications an appearance
of the long-range superconducting order has not been confirmed (see, e.g.,
Ref.~\cite{Aimi07}). Therefore, analytical studies are desirable to elucidate this
problem.
\par
An accurate analytical method is based on the HO technique where the HO algebra is
implemented rigorously. The superconducting pairing induced by the kinematic interaction
for the HOs  was first proposed  by Zaitsev and Ivanov~\cite{Zaitsev87} who studied the
two-particle vertex equation by applying the diagram technique for HOs. The
momentum-independent $s$-wave pairing was found in the lowest order diagram approximation
equivalent to the MFA. However, this solution violates the HO kinematics and  the
$t$--$J$ model should be used to obtain the $d$-wave pairing mediated by the AF
superexchange interaction (see, e.g.,
Refs.~\cite{Plakida89,Yushankhai91,Plakida99,Prelovsek05}). In this respect we should
point out that in many publications superconductivity in  the $t$--$J$ model was studied
in the MFA (see, e.g., Refs.~\cite{Plakida89,Yushankhai91,Valkov02,Jedrak10}). As we have
shown in Sec.~\ref{sec:9}, the intersite Coulomb repulsion suppresses or even destroy
superconductivity induced by the AF superexchange interaction in the MFA. In particular,
in cuprates,  a sufficiently strong n.n. hole repulsion   $V_1 = 0.1-0.2
$~\cite{Feiner96} may be detrimental for the RVB state~\cite{Anderson87}. The same remark
refers to the studies of superconductivity in the conventional Hubbard model in the MFA
(see, e.g., Refs.~\cite{Beenen95,Avella97,Stanescu00,Adam07}). Therefore, consideration
of the spin-fluctuation pairing beyond the MFA is essential  in description of
superconductivity in cuprates as discussed in detail in Sec.~\ref{sec:9}.
\par
In comparison with studies of the intersite Coulomb repulsion in the weak correlation
limit in Refs.~\cite{Alexandrov11,Raghu12}, in the strong correlation limit  the
intersite Coulomb repulsion $V_{i j} $ to some extent is compensated by the nonretarded
superexchange interaction $J_{i j}$ (see Eqs.~(\ref{19a}), (\ref{19b})) which is absent
in the  weak correlation limit. At the same time, even for a sufficiently  large
component $V_{c0}$ of the CI $V_{i j}$, the contribution to the gap equation is given by
a much smaller $d$-wave partial harmonic (\ref{58}) and therefore is not so detrimental
to superconductivity in comparison with the conventional $s$-wave momentum-independent
pairing.
\par
Studies of  the spin-fluctuation $d$-wave pairing in the presence of the EPI have shown
that depending on the symmetry,  the EPI could enhance or suppress superconducting
pairing (see, e.g., Ref.~\cite{Plakida94,Shneider09}). In Ref.~\cite{Lichtenstein95} the
$d$-wave pairing induced by both the spin-fluctuations and EPI in the model (\ref{53})
within the FLEX approximation was considered. It was  revealed that a
momentum-independent EPI strongly suppresses $T_c$, while the EPI with strong forward
scattering can enhance $T_c$.  In our theory the strong spin-fluctuation pairing is
induced by the kinematic interaction which is absent in the weak correlation limit as in
FLEX approximation, and therefore, the EPI plays only a secondary role in the $d$-wave
pairing. A strong EPI in polaronic effects observed in the oxygen isotope effect on the
in-plane penetration depth in cuprates~\cite{Khasanov04} may be irrelevant for the
pairing mediated  by the $d$-wave partial harmonic of the EPI ~\cite{Plakida11}  as
confirmed by a weak isotope effect on $T_c$ in the optimally doped cuprates.

\section{Conclusion}
\label{sec:10}

In the paper the theory of superconducting pairing within the extended  Hubbard model
(\ref{2}) in the limit of strong electron correlations  is presented. Using the Mori-type
projection technique we  obtained a self-consistent system of equations for normal and
anomalous (pair) GFs and for the self-energy calculated in the NCA. The theory  is
similar to the Migdal-Eliasberg strong-coupling approximation.
\par
We can draw the following  conclusions about the mechanism of pairing in the extended
Hubbard model. Solution of the gap equation  in the weak coupling approximation
(\ref{55}) shows that for the $d$-wave pairing the intersite Coulomb repulsion gives a
small contribution determined by $\,l=2\, $ harmonic of the  interaction function $V({\bf
k -q})$. However, it can be larger than the AF superexchange interaction $\, J({\bf k
-q}) \,$,  and the RVB-type superconducting pairing can be destroyed.
\par
Pairing induced by charge fluctuations $\chi_{ch}({\bf k -q})$ appears to be quite weak
(outside the charge-instability region). We have found that  the $d$-wave component of
the EPI, even for the model of strong forward scattering~\cite{Zeyher96} and a large
fully symmetric $s$-wave component, turned out to be small. The largest contribution to
the $d$-wave pairing comes from the electron interaction with spin fluctuations induced
by  strong kinematic interaction $|t({\bf q})|^2$, so that the EPI plays a secondary role
in achieving high-$T_c$.
\par
It is important to point out that the superconducting pairing induced by the AF
superexchange interaction and spin-fluctuation scattering are caused  by the  kinematic
interaction characteristic of systems with strong correlations. These mechanisms of
superconducting pairing are absent in the fermionic models (for a discussion, see
Ref.~\cite{Anderson97}) and are generic for  cuprates. The intersite Coulomb repulsion is
not strong enough to destroy the $d$-wave pairing mediated by spin fluctuations.
Therefore, we believe that the magnetic mechanism of superconducting pairing in the
Hubbard model in the limit of strong correlations is a relevant mechanism of
high-temperature superconductivity in the copper-oxide materials.

{\it Note added in proof.} -- When this work was submitted, we became aware of references
~\cite{Plekhanov03}--\cite{Senechal12} which consider the extended Hubbard model with the
intersite Coulomb repulsion $V$. The results of the references
~\cite{Plekhanov03},~\cite{Senechal12} show that the on-site repulsion $U$ effectively
enhances the $d$-wave pairing which survives for large values of $V$ up  to $V \sim U/2
\gg J $ (Ref.~\cite{Senechal12}). This observation supports our model of spin-fluctuation
pairing due to the kinematic interaction which emerges only in the strong correlation
limit. As long as the Coulomb repulsion $V$ does not exceed  the  kinematic interaction
of the order of the kinetic energy, $ V \lesssim 4 \, t\,$, the $d$-wave pairing may
survive. The small value of $ V = J $ found in Ref.~\cite{Raghu12a} is explained by
application  of the slave-boson representation in the MFA which ignores  the kinematic
interaction. We would like to thank  A.-M. S. Tremblay for valuable discussion who drew
our attention to those  papers.

\acknowledgments

The authors would like to thank A.S. Alexandrov and V.V. Kabanov for valuable
discussions.  One of the authors (N.P.) is grateful to the MPIPKS, Dresden, for the
hospitality during his stay at the Institute, where a part of the present work has been
done. Partial financial support by the Heisenberg--Landau Program of JINR is
acknowledged.

\appendix

\section{Pair correlation function in  MFA}
\label{sec:A}

Here we  calculate  the pair correlation function $\langle X_i^{02} N_{j} \rangle$
considering an equation of motion for the commutator GF $\, L_{ij}(t-t') = \langle
\langle X_{i}^{02} (t) \mid N_j (t') \rangle \rangle|_{i\neq j} \, $. The equation for
the GF can be written as~\cite{Plakida03}:
\begin{eqnarray}
\left( \omega - \varepsilon\sb{2} \right) L\sb{ij}(\omega) & = &
     \sum\sb{m\neq i,\sigma'}  \sigma' \, t\sp{12}\sb{im}
    \{ \langle\langle X\sb{i}\sp{0\bar\sigma'}
     X\sb{m}\sp{0\sigma'} | N\sb{j} \rangle\rangle\sb{\omega}
 \nonumber\\
 &- &\langle\langle X\sb{i}\sp{\sigma' 2}
 X\sb{m}\sp{\bar\sigma' 2} | N\sb{j} \rangle\rangle\sb{\omega} \} ,
\label{A1}
\end{eqnarray}
where we  neglected excitation energy  proportional to the intraband hopping in
comparison with the interband contribution, $|t\sp{\alpha\alpha}\sb{im}| \ll
|\varepsilon\sb{2}| \simeq U \,$. The pair correlation function is determined by the
equation:
\begin{equation}
  \langle X_{i}^{02} N_j \rangle = -\frac{1}{\pi}\int^{+\infty}_{-\infty} \,
      \frac{d\omega}{1- \exp(-\omega/ T)}\,
 {\rm Im}\,  L\sb{ij}(\omega).
 \label{A2}
\end{equation}
The GF $L_{ij}(\omega)$ has two poles, one at $\omega = \varepsilon\sb{2}$ and another at
the energy of a pair excitation given by the GFs at the right-hand side in Eq.~(\ref{A1})
of the singly  or the doubly occupied subbands. Let us consider the hole doped case, $n =
1+ \delta > 1$ when the chemical potential crosses the two-hole subband, $\, \mu \sim
U_{}\,$ and $\varepsilon\sb{1} = -\mu \sim - U_{}, \; \varepsilon\sb{2} =
2\varepsilon\sb{1} + U_{}  \sim - U_{}\,$. In this case  we can neglect the exponentially
small contribution of the order of $\, \exp(- U_{}/T) \ll 1\,$ coming from the pole
$\omega = \varepsilon\sb{2}$. The contribution from the GF of the one-hole subband
in~(\ref{A1}),
\begin{equation}
-\frac{1}{\pi} {\rm Im}
 \langle\langle X_{i}^{0\bar\sigma} X_{m}^{0\sigma} | N_j
 \rangle\rangle_{\omega} \simeq \delta_{mj} \langle
 X_{i}^{0\bar\sigma}X_{j}^{0\sigma} \rangle
 \delta(\omega - 2 \varepsilon_1),
 \label{A3}
\end{equation}
also gives an exponentially small contribution of the order of $\, {\exp }(- 2 U_{} /T)
\ll 1 \,$. Therefore, we  can  take into account only the contribution from the GF $\,
\langle\langle X\sb{i}\sp{\sigma' 2} X\sb{m}\sp{\bar\sigma' 2} | N\sb{j}
\rangle\rangle\sb{\omega}\,$ where the pair excitation energy $\omega \sim
|t\sp{22}\sb{im}|$. Using the approximation $\,1/(\omega - \varepsilon\sb{2} )\simeq
1/U_{} \,$ which neglects retardation effects an integration over $\omega$ in
Eq.~(\ref{A3})  gives the following result,
\begin{eqnarray}
  \langle X_{i}^{02} N_j \rangle &=&
  - \frac{1}{ U_{}}\, \sum_{m\neq i,\sigma'} \, \sigma' \, t^{12}_{im}
  \langle X_{i}^{\sigma' 2} X_{m}^{\bar\sigma' 2}  N_j \rangle
\nonumber\\
 & \simeq &
   - ({4 t^{12}_{ij} }/{ U_{}})  \sigma \,
  \langle X_{i}^{\sigma 2} \, X_{j}^{\bar\sigma 2}\rangle .
 \label{A4}
\end{eqnarray}
The last formula is obtained in the two-site approximation usually applied for the
$t$-$J$ model: $\, m = j $, which gives $X_{j}^{\bar\sigma 2} N_j = 2 X_{j}^{\bar\sigma
2}$.

\section{Self-energy}
\label{sec:B}

The normal and anomalous  (pair) components of the self-energy operator (\ref{23}) are
given by the matrices:
\begin{eqnarray}
 \hat{M}_{ij\sigma}(\omega)=
\langle\! \langle
    { [\widetilde{X_{i}^{\sigma2}}, H] \choose
     [\widetilde{X_{i}^{0 \bar\sigma }}, H]}  \mid
 ( [H, \widetilde{X_{j}^{2\sigma}}]\,
  [H, \widetilde{X_{j}^{\bar\sigma 0}}])
 \rangle \! \rangle^{(pp)}_{\omega}, \quad
\label{B1} \\
\hat{\Phi}_{ij\sigma}(\omega)= \langle\! \langle
    { \widetilde{X_{i}^{\sigma 2}}, H]   \choose
     [\widetilde{X_{i}^{0 \bar\sigma}}, H]} \mid
 ( [H, \widetilde{X_{j}^{\bar\sigma 2}}]
  [H, \widetilde{X_{j}^{0 \sigma}}])
 \rangle \! \rangle^{(pp)}_{\omega}, \quad
 \label{B2}
 \end{eqnarray}
where  $\, [ \widetilde{X_{i}^{\alpha, \beta}} , H] \,$ and $\, [H,
\widetilde{X_{i}^{\alpha, \beta}} ]  \,$ are the  irreducible parts of the commutators
determined by Eq.~(\ref{11}). Using equations of motion for the HOs as given, e.g., by
Eq.~(\ref{7}) we obtain multiparticle GFs which are determined by products of bosonic and
fermionic operators. Let us consider, in particular, contributions to the two-hole
subband self-energy given by the kinematic interaction in Eq.~(\ref{7}). The normal
component in Eq.~(\ref{B1}) reads,
\begin{eqnarray}
&& M^{22 (k) }_{ij\sigma}(\omega) = \sum\sb{l l'\sigma' \sigma''}
 t_{il}t_{jl'} \,\langle\!\langle
B\sb{i\sigma\sigma'} X_l^{\sigma' 2}
 | X_{l'}^{2\sigma''} B\sb{j\sigma\sigma''}^\dag
\rangle\!\rangle_{\omega}
 \nonumber \\
 && =
\sum\sb{l l'\sigma' \sigma''}
 t_{il} t_{jl'} \, \frac{1}{2\pi}\int_{-\infty}^{\infty}
 dz \frac{e^{\beta z}+1}
  {(\omega - z)}\int_{-\infty}^{\infty}\! dt e^{izt}
 \label{B3} \\
&& \times\;\langle X_{l'}^{2\sigma''} B\sb{j\sigma\sigma''}^\dag
 |B\sb{i\sigma\sigma'}(t)
X_l^{\sigma' 2}(t)\rangle .
 \nonumber
\end{eqnarray}
For the anomalous component in Eq.~(\ref{B2}) we have
\begin{eqnarray}
&& \Phi^{22 (k) }_{ij\sigma}(\omega) =  -\sum\sb{l l'\sigma' \sigma''}
 t_{il}t_{jl'} \,\langle\!\langle
B\sb{i\sigma\sigma'} X_l^{\sigma' 2}
 | X_{l'}^{\bar{\sigma}'' 2} B\sb{j\bar{\sigma}\bar{\sigma}''}
\rangle\!\rangle_{\omega}
 \nonumber \\
 && =
- \sum\sb{l l'\sigma' \sigma''}
 t_{il} t_{jl'} \, \frac{1}{2\pi}\int_{-\infty}^{\infty}
 dz \frac{e^{\beta z}+1}{\omega - z}\int_{-\infty}^{\infty}\! dt e^{izt}
 \label{B4} \\
&&\times \;\langle  X_{l'}^{\bar{\sigma}'' 2} B\sb{j \bar{\sigma}\bar{\sigma}''}
 | B\sb{i\sigma \sigma'}(t) X_l^{\sigma' 2}(t)\rangle ,
 \nonumber
\end{eqnarray}
where the bosonic operator $B\sb{i\sigma\sigma'} = B^{22}\sb{i\sigma\sigma'}$ is defined
by Eq.~(\ref{8a}). Using the spectral representation for the thermodynamic
GFs~\cite{Zubarev60} we introduced  in Eqs.~(\ref{B3}),~(\ref{B4}) the multi-particle
time-dependent correlation functions. They are calculated  in the mode-coupling
approximation as described by Eqs.~(\ref{B5}),~(\ref{B6}). The time-dependent
single-particle fermionic and bosonic correlation functions which appear after the
two-time decoupling  are calculated self-consistently as e.g.,
\begin{eqnarray}
\langle   X_{l'}^{2\sigma} X_l^{\sigma 2}(t)\rangle & =& \int_{-\infty}^{\infty} d\omega'
n(\omega') e^{-i\omega' t}
 \nonumber \\
 & \times & [-(1/\pi)]\,{\rm Im} G^{22}_{ll'\sigma}(\omega'),
 \label{B7}\\
\langle B\sb{j\sigma\sigma'}^\dag
 |B\sb{i\sigma\sigma'}(t) \rangle
 & = & \int_{-\infty}^{\infty} d\omega'  N(\omega')
 e^{-i\omega' t}
 \nonumber \\
 & \times & [-(1/\pi)]\,{\rm Im} \langle\! \langle
B\sb{i\sigma\sigma'}|B\sb{j\sigma\sigma'}^\dag
 \rangle \! \rangle_{\omega'}. \quad
 \label{B8}
\end{eqnarray}
Here $G^{22}_{ll'\sigma}(\omega')$ is the GF (\ref{24}) for  the two-hole subband  and
$\langle\! \langle B\sb{i\sigma\sigma'}|B\sb{j\sigma\sigma'}^\dag
 \rangle \! \rangle_{\omega'}$ is the commutator
GF for bosonic excitations. Integration over time $t$ in Eqs.~(\ref{B3}) and (\ref{B4})
yields
\begin{eqnarray}
&& M^{22 (k) }_{ij\sigma}(\omega) = \int \! \int_{-\infty}^{\infty} \frac{d\omega_1
d\omega_2}{\pi^2}
 \; \frac{1 - n(\omega_1)+ N(\omega_2)}
  {\omega -\omega_1 - \omega_2}
\nonumber\\
& &\times  \sum\sb{l l'\sigma'}
 t_{il}t_{jl'}\,{\rm Im} G^{22}_{ll'\sigma'}(\omega_1)\,
 {\rm Im} \langle\! \langle
B\sb{i\sigma\sigma'}|B\sb{j\sigma\sigma'}^\dag
 \rangle \! \rangle_{\omega_2},
  \label{B9}\\
  && \Phi^{22 (k) }_{ij\sigma}(\omega) =
- \int \!\int_{-\infty}^{\infty} \frac{d\omega_1 d\omega_2}{\pi^2} \;  \frac{1 -
n(\omega_1)+ N(\omega_2)}
  {\omega -\omega_1 - \omega_2}
\nonumber\\
&& \times   \sum\sb{l l'\sigma'}
 t_{il}t_{jl'}\,{\rm Im} F^{22}_{ll'\sigma'}(\omega_1)\,
 {\rm Im} \langle\! \langle
B\sb{i\sigma\sigma'}| B\sb{j\bar{\sigma}\bar{\sigma}'}
 \rangle \! \rangle_{\omega_2}.
  \label{B10}
\end{eqnarray}
Taking into account the definition of the bosonic operator (\ref{8a}) the bosonic GFs in
these equations can be written as
\begin{eqnarray}
&&\langle\! \langle B\sb{i\sigma\sigma'}| B\sb{j\sigma\sigma'}^\dag
 \rangle \! \rangle_{\omega}
=  (1/4)\langle\! \langle N_{i} | N_{j}\rangle \! \rangle_{\omega} \,
\delta_{\sigma'\sigma}
 \nonumber \\
&&  + \langle\! \langle  S^z_{i}| S^z_{j}
 \rangle \!\rangle_{\omega} \, \delta_{\sigma'\sigma}
 + \langle\! \langle X^{\bar{\sigma}\sigma}_{i}|
 X^{\sigma \bar{\sigma}}_{j}\rangle \! \rangle_{\omega}
 \, \delta_{\sigma' \bar{\sigma}},
 \label{B11}\\
 &&\langle\! \langle B\sb{i\sigma\sigma'}|
 B\sb{j\bar{\sigma}\bar{\sigma}'}
 \rangle \! \rangle_{\omega}
=  (1/4)\langle\! \langle N_{i} | N_{j}\rangle \! \rangle_{\omega} \,
\delta_{\sigma'\sigma}
 \nonumber \\
&&  - \langle\! \langle  S^z_{i}| S^z_{j}
 \rangle \!\rangle_{\omega} \, \delta_{\sigma'\sigma}
 + \langle\! \langle X^{\bar{\sigma}\sigma}_{i}|
 X^{\sigma \bar{\sigma}}_{j}\rangle \! \rangle_{\omega}
 \, \delta_{\sigma' \bar{\sigma}}.
  \label{B12}
\end{eqnarray}
After summation over $\sigma'$ in (\ref{B9}) for the bosonic GF (\ref{B11}) and the
normal GF in the paramagnetic state, $ G^{22}_{ll'\sigma}(\omega) = G^{22}_{ll' \bar
\sigma}(\omega)$, the spin-fluctuation contribution  can be written in the form:
$\langle\! \langle  S^z_{i}| S^z_{j}
 \rangle \!\rangle_{\omega}
 + \langle\! \langle X^{\bar{\sigma}\sigma}_{i}|
 X^{\sigma \bar{\sigma}}_{j}\rangle \! \rangle_{\omega} =
 \langle\! \langle {\bf S}_{i}|{\bf S}_{j}
 \rangle \!\rangle_{\omega} $. Similar summation over $\sigma'$ in
(\ref{B10}) for the bosonic GF (\ref{B12}) and the anomalous GF $
F^{22}_{ll'\sigma}(\omega) = - F^{22}_{ll' \bar \sigma}(\omega)$, results in the
equation: $\, - \langle\! \langle S^z_{i}| S^z_{j} \rangle\!\rangle_{\omega}\,
 F^{22}_{ll'\sigma}(\omega)
  + \langle\! \langle X^{\bar{\sigma}\sigma}_{i}|
 X^{\sigma \bar{\sigma}}_{j}\rangle \! \rangle_{\omega} \,
 F^{22}_{ll' \bar \sigma}(\omega)  = -
 \langle\! \langle {\bf S}_{i}|{\bf S}_{j}
 \rangle \!\rangle_{\omega}\, F^{22}_{ll'\sigma}(\omega) $.

Introducing the $\bf q$-representation for the GFs and the self-energies as defined by
Eq. (\ref{9fk})   for the self-energies (\ref{B9}) and (\ref{B10}) we obtain the
expressions:
\begin{eqnarray}
 M^{22 (k) }({\bf k},\omega) &= &
   \frac{1}{N} \sum\sb{\bf q}
   \int\limits\sb{-\infty}\sp{+\infty} \!\!{\rm d}z\,
   K^{(+)}_{(k)}(\omega,z|{\bf q },{\bf k-q})
   \nonumber \\
& \times & [ - ({1}/{\pi})\, \mbox{Im}\,
  G\sp{22}({\bf q},z)] ,
 \label{B13}
\end{eqnarray}
\begin{eqnarray}
\Phi\sb{\sigma}^{22 (k) }({\bf k},\omega) & = &
 \frac{1}{N} \sum\sb{\bf q}
   \int\limits\sb{-\infty}\sp{+\infty} \!\!{\rm d}z\,
   K^{(-)}_{(k)}(\omega,z|{\bf q },{\bf k-q})
  \nonumber \\
& \times & [- ({1}/{\pi}) \,\mbox{Im}\,
 F^{22}\sb{\sigma}({\bf q},z)] ,
 \label{B14}
\end{eqnarray}
where the contribution from the kinematic interaction  is given by the kernel
\begin{eqnarray}
 &&  K^{(\pm)}_{(k)}(\omega,z |{\bf q },{\bf k -q })  =
\frac{|t({\bf q})|^{2}} {\pi} \int\limits\sb{-\infty}\sp{+\infty} d \Omega \frac{1
+N(\Omega) - n(z)}{\omega - z - \Omega}
   \nonumber \\
&&\times \big\{ {\rm Im}\, \chi\sb{sf}({\bf k- q},\Omega) \pm(1/4) {\rm Im}\,
\chi_{cf}({\bf k-q}, \Omega) \big\} .
  \label{B15}
\end{eqnarray}
Here the spin- and charge-susceptibility are defined by Eqs.~(\ref{32a}) and (\ref{32b}).
By taking into account contributions from the CI and the EPI in Eq.(\ref{7}) and using
the NCA in  calculation of the respective time-dependent correlation functions we obtain
the kernel (\ref{31}) for  the integral equations (\ref{29}) and (\ref{30}).


\begin{thebibliography}{99}
\bibitem{Bednorz86}  J.\ G. Bednorz and K.\ A. M\"{u}ller,
Z. Phys. B. \textbf{64}, 189 (1986).
% Possible high Tc superconductivity in the Ba-La-Cu-O system.
\bibitem{Schrieffer07}  \textit{ Handbook of High-Temperature  Superconductivity.
Theory and Experiment}, edited by J.R.~Schrieffer  and J.S.~Brooks (Springer-Verlag, New
York, 2007), Chaps. 13--15.
\bibitem{Plakida10}  N.\ M. Plakida,
\textit{ High-Temperature Cuprate Superconductors} (Springer Series in Solid-State
Sciences, Vol. 166, Springer-Verlag, Berlin, 2010), Chap. 7.
\bibitem{Anderson87}  P.\ W.~Anderson,  Science
\textbf{235},  1196 (1987);  P.\ W.~Anderson, \textit{The theory of superconductivity in
the high-$T\sb{c}$ cuprates} (Princeton University Press, Princeton, 1997).
\bibitem{Scalapino95}  D.J. Scalapino,  Phys.  Reports
{\bf 250}, 329 (1995).
% [arXiv:cond-mat/9908287].  The case for
%$d_{x^2 -y^2}$ pairing in the cuprate superconductors.
\bibitem{Monthoux94b} P. Monthoux and D. Pines,
 Phys. Rev. B {\bf 49}, 4261 (1994).
% Spin-fluctuation-induced  superconductivity and
% normal-state properties of  YBa$_2$Cu$_3$O$_{7}$.
\bibitem{Moriya00} T. Moriya and K. Ueda,
Adv. in Physics  {\bf 49}, 555  (2000);
% Spin fluctuations and high temeperature superconductivity.
 Rep. Prog. Phys.  {\bf 66}, 1299 (2003).
% Antiferropmagnetic spin fluctuation and  superconductivity.
\bibitem{Chubukov04}  A. V. Chubukov, D. Pines, and J. Schmalian,
in {\it The Physics of Conventional and Unconventional Superconductors}, edited by K. H.
Bennemann and J. B. Ketterson (Springer-Verlag, Berlin, 2004), Vol. I, p. 495;
%  [arXiv:cond-mat/0201140]. A spin fluctuation
%model for $d$-wave  superconductivity.
 Ar.~Abanov, A.V. Chubukov,  and J. Schmalian, Advances in
Phys. {\bf 52}, 119 (2003).
%  Quantum-critical theory of spin-fermion model and its
% application to cuprates: normal state analysis.
\bibitem{Abanov08} Ar. Abanov, A.V. Chubukov, and M.R. Norman,
Phys. Rev. B \textbf{78}, 220507(R) (2008).
% Gap anisotropy and universal pairing scale in a spin-fluctuation
% model of cuprate superconductors
\bibitem{Kordyuk10}   A.A. Kordyuk, V.B. Zabolotnyy, D.V. Evtushinsky, D.S. Inosov, T.K. Kim,
B. B\"{u}chner, and S.V. Borisenko,  Eur. Phys. J. \textbf{188}, 153 (2010).
%An ARPES view on the high-$T_c$ problem: Phonons vs. spin-fluctuations
\bibitem{Dahm09} T.~Dahm, V. Hinkov, S.V. Borisenko, A.A. Kordyuk,
 V.B. Zabolotnyy, J. Fink, B. B\"{u}chner, D.J. Scalapino,
 W. Hanke, and  B. Keimer,  Nature Phys. \textbf{5}, 780 (2009).
%Strength of the spin-fluctuation mediated pairing interaction in a
%high-temperature superconductor.
\bibitem{Bourges98} Ph. Bourges, in  {\it The Gap Symmetry
and Fluctuations in High Temperature  Superconductors},
 edited by J.~Bok, G.~Deutscher, D.~Pavuna and  S.\,A.~Wolf
(Plenum Press, 1998), p. 349.
%(Vol. \textbf{371} in NATO ASI series, Physics). [arXiv:cond-mat/9901333].
%From magnons to the resonance peak: Spin dynamics in high-$T_c$ superconducting cuprates by
%inelastic neutron scattering.
\bibitem{LeTacon11} M. Le Tacon,
G. Ghiringhelli, J. Chaloupka, M. Moretti Sala, V. Hinkov, M.W. Haverkort, M. Minola, M.
Bakr, K. J. Zhou, S. Blanco-Canosa,
 C. Monney, Y. T. Song, G. L. Sun, C. T. Lin, G. M. De Luca,
 M. Salluzzo, G. Khaliullin, T. Schmitt, L. Braicovich, and B. Keimer,
 Nature Phys. \textbf{7}, 725  (2011).
%Intense paramagnon excitations in a large family of
%high-temperature superconductors.
\bibitem{Vladimirov12} A.A. Vladimirov, D. Ihle, and N.M. Plakida,
Phys. Rev. B \textbf{85}, 224536  (2012).
% Optical and dc conductivities of  cuprates:
% Spin-fluctuation scattering in the t-J model,
\bibitem{Kulic04}  M.\ L. Kuli\'{c}, in {\it Lectures on Physics
of Highly Correlated Electronic Systems VIII}, edited by A.~Avella and F.~Mancini, AIP
Conf. Proc.,  Vol. {715} (Melville, New York, 2004), p. 75.
% doi:http://dx.doi.org/10.1063/1.1800734.
% Electron- Phonon Interaction and Strong Correlations
%in High-Temperature Superconductors:
% One can not avoid the unavoidable. arXiv: cond-mat/0404287
\bibitem{Maksimov10}  E.G. Maksimov,  M.L. Kuli\'{c}, and O.\ V. Dolgov, Adv. in
Cond. Mat. Phys., Volume 2010, Article ID 423725
 (2010) ( DOI: 10.1155/2010/423725)
%Bosonic Spectral Function and the Electron-Phonon Interaction in
%HTSC Cuprates
\bibitem{Raghu10} S. Raghu, S. A. Kivelson, and D.J. Scalapino,
 Phys. Rev. B {\bf 81}, 224505 (2010).
% Superconductivity in the repulsive Hubbard model:
% An asymptotically exact weak-coupling solution.
\bibitem{Hubbard63} J. Hubbard, Proc. Roy. Soc. (London) A,
\textbf{276},  238 (1963).
%Electron correlations in narrow energy bands.
\bibitem{Alexandrov11} A.\ S. Alexandrov and V.V. Kabanov,
Phys. Rev. Lett. \textbf{106}, 136403 (2011).
% Unconventional high-temperature superconductivity from repulsive
%interactions: theoretical constraints
\bibitem{KaganM11} M.\ Yu. Kagan, D.V. Efremov, M.S. Marienko, and V.S. Valkov.
JETP Lett. \textbf{93} 725 (2011).
%Triplet p-Wave Superconductivity in the Low-Density Extended Hubbard Model with Coulomb Repulsion.
\bibitem{Efremov00} D.\ V. Efremov, M.\ S. Mar’enko, M.A. Baranov, and M.Yu. Kagan,
J. Exp. Theor. Phys.  \textbf{90}, 861 (2000).
\bibitem{Raghu12} S. Raghu,   E. Berg, A.V. Chubukov,
and S.\ A. Kivelson1  Phys. Rev. B {\bf 85}, 024516 (2012).
%%  Effects of longer-range interactions on unconventional
%superconductivity:  Strong correlations
\bibitem{Fulde95}  P. Fulde, {\it Electronic correlations
in molecules  and solids} (Springer Verlag, Berlin 1995).
% 3rd ed., 480 pp.
\bibitem{Avella11} {\it  Strongly
Correlated Systems. Theoretical Methods}, edited by A.~Avella and F.~Mancini (Springer
Series in Solid-State Sciences, Vol. 171, Springer Verlag, Berlin, 2012).
\bibitem{Zeyher96} R. Zeyher and  M. L. Kuli\'{c},
 Phys. Rev. B {\bf 53}, 2850 (1996).
% Renormalization of the electron-phonon interaction by strong
%electronic correlations in high-Tc superconductors
\bibitem{Mori65} H. Mori, Prog. Theor. Phys. \textbf{34},
399 (1965).
% A continued-fraction representation of the time-correlation
%functions.
\bibitem{Plakida07}   N.M.~Plakida  and  V.S.~Oudovenko,
JETP \textbf{104}, 230 (2007).
%Electronic spectrum in high-temperature cuprate superconductors.
\bibitem{Hubbard65} J. Hubbard,
Proc.~Roy. Soc. A (London,) \textbf{285}, 542 (1965).
% Electron correlations in narrow energy bands.  IV.  The atomic
% representation.
\bibitem{Feiner96}  L.F. Feiner, J.H. Jefferson, and  R.~Raimondi,
{ Phys.~Rev.~B} \textbf{53}, 8751 (1996).
% Effective single-band models for high-$T_c$ cuprates. I. Coulomb
%interactions
\bibitem{Emery87} V.J. Emery,  Phys.~Rev.~Lett.  \textbf{58},
  2794 (1987);
% Theory of high-$T_c$ in oxides,
  C.M.~Varma, S.~Schmitt-Rink, and E.~Abrahams,
Solid State Commun.  \textbf{62}, 681 (1987).
% Charge transfer excitations and superconductivity in ionic
% metals,
 \bibitem{Zhang88} F.C. Zhang and  T.M.~Rice,
  Phys.~Rev.~B  \textbf{37}, 3759  (1988).
%Effective Hamiltonian for the superconducting Cu oxides
\bibitem{Zubarev60} D.N. Zubarev,
 %Double-time Green's functions in statistical  physics,
{ Usp. Fiz. Nauk}  \textbf{71}, 71  (1960); ({ Sov. Phys. Usp.} \textbf{3}, 320 (1960));
{\it Nonequilibrium Statical Thermodynamics} (Consultant Bureau, New-York, 1974).
\bibitem{Adam07} Gh.\, Adam  and S.\, Adam,
J.~Phys. A:  Math. Theor. {\bf 40}, 11205 (2007).
% Rigorous derivation of the mean  field Green functions of
%the two-band Hubbard model of  superconductivity.
\bibitem{Plakida03} N.M. Plakida,   L.~Anton, S.~Adam,
and Gh.~Adam, Zh. Exp.Theor. Fyz. \textbf{124}, 367 (2003), (JETP \textbf{97}, 331
(2003)).
% Exchange and spin-fluctuation superconducting pairing
%in the Hubbard model in the strong correlation limit.
\bibitem{Plakida97} N.M. Plakida, Physica C \textbf{282--287}, 1737 (1997).
%Superconductivity in the two-band singlet-hole model for
%copper-oxide plane
\bibitem{Migdal58} A.B.~Migdal, Zh. Eksp.  Teor. Fiz. \textbf{34}, 1438 (1956),
 (Soviet~Phys.~JETP  \textbf{7}, 996 (1958)).
 \bibitem{Eliashberg60} G.M.~Eliashberg, Zh. Eksp. Teor. Fiz. \textbf{38}, 966 (1960);
ibid  \textbf{39}, 1437 (1960) (Soviet Phys.~JETP \textbf{11}, 696 (1960); ibid
\textbf{12},  1000 (1960)).
\bibitem{Liu92} Z.~Liu and E.~Manousakis,
Phys. Rev. B \textbf{45}, 2425 (1992).
%Dynamical properties of a hole in a Heisenberg antiferromagnet
\bibitem{Monthoux97} P. Monthoux,
Phys. Rev. B {\bf 55}, 15261 (1997).
% Vertex corrections and two-loop pairing potential
%in nearly antiferromagnetic Fermi liquids.
\bibitem{Becca96} F. Becca, M. Tarquini, M. Grilli,
and C. Di Castro, Phys.~Rev.~B, \textbf{54},  12 443 (1996).
 %  Charge-density waves and superconductivity as an alternative
 % to phase separation in the infinite-U Hubbard-Holstein model
\bibitem{Castellani98} C. Castellani, C.  Di~Castro,  and M. Grilli, J. of Phys. and Chem. of Sol. {\bf 59},
1694 (1998).
% Stripe formation:  A quantum critical point
%for  cuprate superconductors.
\bibitem{Gabovich11} T. Ekino, A.M. Gabovich, Mai Suan Li,
M. P\c{e}ka\l a, H.~Szymczak, and A.I.~Voitenko, J. Phys.: Condens. Matter \textbf{23}
385701 (2011).
%(22pp) The phase diagram for coexisting d-wave
% superconductivity and charge-density waves: cuprates and beyond
\bibitem{Jaklic95} J. Jakli\v{c} and P. Prelov\'sek,
 Phys. Rev. Lett. \textbf{74}, 3411 (1995);
{ ibid.} \textbf{75}, 1340 (1995).
\bibitem{Vladimirov09} A.A. Vladimirov, D. Ihle, and N. M. Plakida,
Phys. Rev. B \textbf{80}, 104425 (2009).
\bibitem{Lichtenstein95}  A.I. Lichtenstein. and  M.L. Kuli\'{c},
  Physica C  {\bf 245}, 186 (1995).
% Electron-boson interaction can  help $d$-wave pairing.
% Self-consistent approach.
\bibitem{Georges96}  A. Georges,  G.~Kotliar, W.~Krauth,
 and M.~Rozenberg, { Rev. Mod. Phys.} \textbf{68}, 13 (1996).
%Dynamical mean-field theory of strongly correlated fermion
%systems and the limit of infinite dimensions.
 \bibitem{Kotliar06} G.~Kotliar, S. Y.~ Savrasov, K.~Haule, V.S.~Oudovenko,
O.~Parcollet, and C.A.~Marianetti,  Rev. Mod. Phys. \textbf{78},
865 (2006).% cond-mat/0511085.
% Electronic structure calculations with dynamical mean-field theory
\bibitem{Haule07} K. Haule and G. Kotliar, Phys. Rev. B
{\bf 76}, 104509 (2007).
% Strongly correlated superconductivity:
% A plaquette dynamical mean-field theory study
\bibitem{Vilk95} Y. Vilk and A.-M. Tremblay, J. Phys. Chem. Solids
(UK) \textbf{56}, 1769 (1995); Y. Vilk and A.-M. Tremblay, J. Phys I (France) \textbf{7},
1309 (1997);
 Y. Vilk, L. Chen, and A.-M. Tremblay,
  Phys. Rev. B  \textbf{49}, 13267 (1994).
%Theory of spin and charge fluctuations in the Hubbard model
\bibitem{Tremblay06} A-M.S.~Tremblay, B.~Kyung and D.~S\'{e}n\'{e}chal,
  Fiz. Nizk. Temp. (Low Temp. Phys., Ukraine)
\textbf{32}, 561 (2006);
% Pseudogap and high-temperature superconductivity from
%weak to strong coupling. Towards quantitative theory
B. Davoudi and A.-M.S. Tremblay,  Phys. Rev. B \textbf{76}, 085115 (2007).
%Non-Perturbative Treatment of Charge and Spin Fluctuations
% in the Two-Dimensional Extended Hubbard Model:
% Extended Two-Particle Self-Consistent Approach
\bibitem{Bickers89}  N.E. Bickers,  D.J. Scalapino,
 and S.R. White, Phys.~Rev.~Lett. {\bf 62}, 961 (1989).
% Conserving approximations for strongly correlated electron
% systems: Bethe-Salpeter equation
%and dynamics for the two-dimensional Hubbard model.
\bibitem{Monthoux94a} P. Monthoux   and  D.J. Scalapino,
Phys. Rev. Lett. {\bf 72}, 1874 (1994).
%  Self-consistent $d_{x^2-y^2}$ pairing in a
% two-dimensional Hubbard model.
\bibitem{Manske04} D. Manske,  I. Eremin,  and
K.H. Bennemann,   in {\it The Physics of Conventional and Unconvencional
Superconductors}, edited by K.H.~Bennemann and J.B.~Ketterson  (Springer-Verlag, Berlin,
2004), Vol.~II, p.~731.
%Electronic Theory for  Superconductivity in  High-$T_c$ Cuprates
%and Sr$_2$RuO$_4$.
% RG and weak coupling
\bibitem{Shankar94}   R. Shankar, Rev. Mod. Phys. \textbf{66}, 129 (1994).
% Renormalization-group approach to interacting fermions.
\bibitem{Metzner98} W. Metzner, C. Castellani, and C. Di Castro,
Adv. Phys. \textbf{47}, 317 (1998).
\bibitem{Kivelson03}  S.A. Kivelson, E. Fradkin,
V. Oganesyan,  I.P. Bindloss, J.M. Tranquada, A. Kapitulnik, and
 C.  Howald, Rev. Mod. Phys. \textbf{75}, 1201 (2003).
 %[arXiv:cond-mat/0210683]   How to detect fluctuating stripes in
%HTSC.
\bibitem{Vojta09} M. Vojta, Adv. Phys., \textbf{58}, 699 (2009).
%Lattice symmetry breaking in cuprate superconductors: stripes,
%nematics, and superconductivity
\bibitem{Furukawa98} N. Furukawa, T.M. Rice, and M. Salmhofer, Phys. Rev. Lett.
\textbf{81}, 3195 (1998).
\bibitem{Honer01} C. Honerkamp, M. Salmhofer, N. Furukawa,
 and T. M. Rice, Phys. Rev. B \textbf{63}, 035109 (2001).
% Breakdown of the Landau-Fermi liquid in two dimensions
% due to umklapp scattering
\bibitem{Honer07} C. Honerkamp, H. C. Fu, and D.-H. Lee,
Phys. Rev. B \textbf{75},  014503 (2007).
%Phonons and d-wave pairing in the two-dimensional Hubbard model
\bibitem{Halb00a} C. J. Halboth and W. Metzner,
Phys. Rev. B \textbf{61}, 7364 (2000).
% Renormalization-group
% analysis of the two-dimensional Hubbard model
\bibitem{Halb00b} C. J. Halboth and W. Metzner,
Phys. Rev. Lett.  \textbf{85}, 5162 (2000)
%d-Wave Superconductivity and Pomeranchuk Instability in the
%Two-Dimensional Hubbard Model
\bibitem{Dagotto94} E. Dagotto, Rev. Mod. Phys. \textbf{66},
 763 (1994).
%Correlated electrons in high-temperature superconductors
\bibitem{Bulut02} N. Bulut, { Advances in Physics}
 \textbf{51}, 1587 (2002).
% $d(x^2-y^2)$ superconductivity and the Hubbard model,
% phenomenological approach
\bibitem{Scalapino07}  D.J. Scalapino,
in \textit{ Handbook of High-Temperature  Superconductivity. Theory and Experiment},
edited by J.R.~Schrieffer  and J.S.~Brooks (Springer-Verlag, New York, 2007), pp.
495--526.
%[arXiv:cond-mat/0610710]. Numerical  studies
% of the 2D Hubbard model.
\bibitem{Maier05} Th. Maier, M.~Jarrel,  Th.~Pruschke,
and M.H.~Hettler,  Rev. Mod. Phys. \textbf{77}, 1027 (2005).
%Quantum cluster theories
\bibitem{Maier06} Th.A. Maier, M. Jarrell,  and
D.J. Scalapino, Phys.  Rev. Lett.  {\bf 96}, 047005 (2006);
%Structure of the  Pairing interaction in the two-dimensional
%Hubbard model.
ibid,  Phys. Rev. B {\bf 74}, 094513 (2006).
% Pairing interaction in the two-dimensional Hubbard model
% studied with a dynamic cluster quantum Monte Carlo approximation
\bibitem{Kancharla08} S. S. Kancharla, B. Kyung, D. S\'{e}n\'{e}chal,
 M. Civelli, M. Capone, G. Kotliar, and A.-M. S. Tremblay,
  Phys. Rev. B  \textbf{77}, 184516 (2008).
% Anomalous superconductivity and its competition
% with antiferromagnetism in doped Mott insulators
\bibitem{Avella07} A. Avella and F. Mancini,
Phys. Rev. B  {\bf 75},  134518 (2007);
% Underdoped cuprate phenomenology in the two-dimensional  Hubbard model
%within the composite operator method.
 A. Avella and F. Mancini, J. Phys.: Condens. Matter {\bf 19},
  255209  (2007).
%The 2D Hubbard model  and the pseudogap: a COM (SCBA) study.
\bibitem{Macridin09} A. Macridin and M. Jarrell,
 Phys. Rev. B {\bf 79}, 104517 (2009).
 % Isotope effect in the Hubbard model with local phonons
\bibitem{Aimi07} T. Aimi  and M. Imada,  J. Phys. Soc. Jpn.
{\bf 76}, 13708 (2007).
% Does simple two-dimensional Hubbard model
% account for high-$T_c$  superconductivity in copper oxides?
\bibitem{Zaitsev87} R.O. Zaitsev,  and  V.A. Ivanov,
{Soviet Phys. Solid State} \textbf{29},    2554  (1987),
% On the possibility of pair condensation in the Hubbard model.
{ Ibid.} {\bf 29},   3111 (1987), {Int. J. Mod. Phys. B} \textbf{5},  153  (1988).
% Superconductivity in the Hubbard model.
\bibitem{Plakida89} N.M. Plakida, V.Yu. Yushankhai, and
I.V. Stasyuk, { Physica C}  \textbf{160}, 80 (1989).
% On the role of kinematical and exchange interactions in
%superconducting pairing of electrons in the Hubbard model,
\bibitem{Yushankhai91} V.Yu. Yushankhai, N.M. Plakida,  and P. Kalinay,
{ Physica C} \textbf{174}, 401 (1991).
% Superconducting pairing in the mean-field
%approximation for the $t-J$ model: Numerical analysis.
\bibitem{Plakida99}   N.M.~Plakida  and  V.S.~Oudovenko,
 { Phys.~Rev.~B}  \textbf{59}, 11949 (1999).
 %Electron spectrum and superconductivity in the t-J model at
%moderate doping.
\bibitem{Prelovsek05}  P. Prelov\v{s}ek and A. Ram\v{s}ak,
 Phys.~Rev.~B {\bf 72}, 012510 (2005).
 % Spin-fluctuation mechanism of  superconductivity in cuprates.
\bibitem{Valkov02} V.V. Val'kov, T.A. Val'kova, D.M. Dzebisashvili,
 and S.G. Ovchinnikov,  JETP Letters  {\bf 75}, 378 (2002).
  %[Pis'ma Zh. Theor. Exper. Phys. {\bf 75}, 450].
% The strong effect  of three-center interactions on the
% formation of  superconductivity with $\,d_{x^2-y^2}\,$
% symmetry in the  $t$--$J^{*}$ model.
\bibitem{Jedrak10} J. J\c{e}drak and J. Spa\l ek,
 Phys. Rev. B \textbf{81}, 073108 (2010),
%Consistent statistical treatment of the renormalized
% mean-field t- J model
ibid., \textbf{83}, 104512 (2011).
% Renormalized mean-field %t- J model of high-Tc
%superconductivity: Comparison to experiment
\bibitem{Beenen95} J. Beenen and D.M. Edwards,
Phys. Rev. B, \textbf{52}, 13636 (1995).
%Superconductivity in the two-dimensional Hubbard model.
\bibitem{Avella97} A.~Avella, F.~Mancini, D.~Villani, and H.~Matsumoto,
Physica C, \textbf{282--287}, 1757 (1997);
%The superconducting gap in the two-dimensional Hubbard model.
T. Di~Matteo, F.~Mancini, H. Matsumoto, and V.S. Oudovenko, Physica B, \textbf{230--232},
915 (1997).
%Singlet pairing in the 2D Hubbard model.
\bibitem{Stanescu00} T.D. Stanescu, I. Martin, and Ph. Phillips,
Phys.~Rev.~B \textbf{62}, 4300 (2000).
%d(x2-y2) pairing of composite excitations in the 2D Hubbard model.
\bibitem{Plakida94}  N.M. Plakida and  R. Hayn,  Z.
Physik B  {\bf 93},  313 (1994).
% Supercoducting pairing in the  singlet band of the Emery model. Supercoducting pairing in the singlet
%band of the Emery model.
\bibitem{Shneider09} E.I. Shneider and S.G. Ovchinnikov,
Zh. Eksp. Teor. Fiz. \textbf{136}, 1177 (2009).
%Изотопический эффект в модели сильно коррелированных
% электронов, учитывающей магнитный и
% фононный механизмы сверхпроводящего спаривания
%Шнейдер Е.И., Овчинников С.Г. ЖЭТФ, 2009 г., Том 136,
% Вып. 6, стр. 1177
\bibitem{Khasanov04}  R. Khasanov, A. Shengelaya,
E. Morenzoni,  K. Conder, I.M. Savi\`{c},  and H. Keller, J. Phys.: Condens. Matter {\bf
16}, S4439 (2004).
%The oxygen isotope effect on the in-plane penetration
%depth in cuprate superconductors
\bibitem{Plakida11}  N.M. Plakida,  Physica Scripta \textbf{83},
   038303 (2011).
%Comments on the paper ``The pairing mechanism
%of high-temperature superconductivity: experimental constraints''
\bibitem{Anderson97}   P.W. Anderson, Adv. in Physics,
\textbf{46}, 3 (1997).
% A re-examination of concepts in magnetic metals:
%the 'nearly antiferromagnetic Fermi liquid'
\bibitem{Plekhanov03} E. Plekhanov, S. Sorella, and M. Fabrizio,
 Phys. Rev. Lett. {\bf 90}, 187004  (2003)
\bibitem{Raghu12a} S. Raghu, R. Thomale, and T. H. Geballe, Phys. Rev. B  {\bf
86}, 094506 (2012)
\bibitem{Senechal12}  D. S\'{e}n\'{e}chal, A. Day, V. Bouliane, and A.-M. S.
Tremblay, arXiv:1212.4503 (unpublished)
\end{thebibliography}
\end{document}